\RequirePackage{fix-cm}

\documentclass[english]{article}

\date{}

\usepackage{graphicx}

\usepackage[utf8]{inputenc}
\usepackage[T1]{fontenc}
\usepackage{bm}
\usepackage{amsmath}
\usepackage{amssymb}
\usepackage{setspace}
\usepackage{natbib}
\usepackage{color}
\usepackage{mathrsfs}
\usepackage{mathtools}
\usepackage{amsfonts}
\usepackage{booktabs,siunitx}
\usepackage[hyphens]{url}
\usepackage[hidelinks]{hyperref}
\hypersetup{breaklinks=true}
\usepackage{tikz}
\usetikzlibrary{decorations.pathmorphing,patterns}
\usetikzlibrary{calc,angles,quotes}
\usetikzlibrary{intersections,arrows}
\usetikzlibrary[shapes,arrows]
\tikzset{ar/.style={->,thick,shorten <=8pt,shorten >=8pt,>=stealth}}
\usepackage{caption} 
\newcommand{\Up}{\mathscr{U}}
\newcommand{\TE}{\mathscr{E}}
\allowdisplaybreaks

\begin{document}

\title{Uniform formulation for orbit computation: \\
  the intermediate elements}

\author{Giulio Ba\`u\footnote{Department of Mathematics, University of
    Pisa, Largo Pontecorvo 5, 56127, Pisa, Italy} \and
        Javier Roa\footnote{Jet Propulsion Laboratory, California Institute of
    Technology, 4800 Oak Grove Drive, Pasadena, CA 91109, USA}}

\maketitle

\begin{abstract}
  We present a new method for computing orbits in the perturbed
  two-body problem: the position and velocity vectors of the
  propagated object in Cartesian coordinates are replaced by eight
  orbital elements, i.e. constants of the unperturbed motion. The
  proposed elements are uniformly valid for any value of the total
  energy. Their definition stems from the idea of applying Sundman's
  time transformation in the framework of the projective decomposition
  of motion, which is the starting point of the Burdet--Ferr{\'a}ndiz
  linearisation, combined with Stumpff's functions. In analogy with
  Deprit's ideal elements, the formulation relies on a special
  reference frame that evolves slowly under the action of external
  perturbations. We call it the \textit{intermediate} frame, hence the
  name of the elements. Two of them are related to the radial motion,
  and the next four, given by Euler parameters, fix the orientation of
  the intermediate frame. The total energy and a time element complete
  the state vector.  All the necessary formulae for extending the
  method to orbit determination and uncertainty propagation are
  provided. For example, the partial derivatives of the position and
  velocity with respect to the intermediate elements are obtained
  explicitly together with the inverse partial derivatives. Numerical
  tests are included to assess the performance of the proposed special
  perturbation method when propagating the orbit of comets C/2003~T4
  (LINEAR) and C/1985~K1 (Machholz).

\end{abstract}



\section{Introduction}
\label{sec:intro}
\citet{Stumpff_1947, Stumpff_1962} devised a method to represent the
solution of the two-body problem at any time $t$ from the position
(${\bf r}_0$) and velocity ($\dot{{\bf r}}_0$) at some reference epoch
$t_0$ \citep[see also][vol.~1, chap.~V]{Stumpff_1959}. His formulation
is very attractive because it is the same for all types of orbits
including those that are rectilinear. For this reason, the coordinates
of ${\bf r}_0$, $\dot{{\bf r}}_0$ are commonly referred to as
\textit{universal variables} (or elements). Stumpff introduced a new
independent variable, here denoted by $\chi$, through the time
transformation (he assumed $\mu=1$)
\begin{equation}
  \frac{\mathrm{d} t}{\mathrm{d} \chi}=\frac{r}{\sqrt\mu},\quad \mu=k^2(m+M),
  \label{eq:chi}
\end{equation}
where $k^2$ is the gravitational constant and $M$, $m$ are the masses
of the two bodies. Equation~(\ref{eq:chi}) together with the Keplerian
energy integral produce a third-order linear differential equation
with constant coefficients for the orbital distance $r$ and the
Lagrangian functions. Then, a unique solution for any type of conic
section can be written in terms of the constants of the motion ${\bf
  r}_0$, $\dot{{\bf r}}_0$ and \textit{Stumpff's functions} $c_n$ (see
Eq.~\ref{eq:cn}). Inserting the solution $r(\chi)$ in the right-hand
side of Eq.~(\ref{eq:chi}) and integrating with the initial condition
$\chi(t_0)=0$, he obtained the generalised form of Kepler's equation,
from which the value of $\chi$ corresponding to $t$ can be determined
by an iterative algorithm.

Samuel Herrick was among the first to get interested in Stumpff's
work. Since the mid 1940s, he devoted his efforts to improving orbit
computation for near parabolic and near rectilinear motion
\citep{Herrick_1945, Herrick_1953}. His universal formulae of the
two-body problem (\citealt[][sect.~6P]{Herrick_1960};
\citealt{Herrick_1965}) are simpler and more convenient than those
proposed by Stumpff. The same formulation was presented by
\citet{Wong_1962} and a similar version by
\citet[][sect.~2.8]{Battin_1964}. Herrick's method relying on
universal variables, which is nowadays regarded as the classic
solution, uses a universal anomaly defined by the differential
relation~(\ref{eq:chi}). \citet{Herrick_1965} showed alternative forms
of the universal variables, by introducing two arbitrary
parameters. One of them allows for a more general definition of the
universal anomaly, given by $\psi=\chi/\sqrt\beta$. Among the six
different choices for $\beta$ that are considered, two deserves
special attention: if we set $\beta=(t-t_0)^2/r_0^2$ then Stumpff's
form is obtained; the choice $\beta=\mu$, proposed by
\citet{Goodyear_1965, Goodyear_1966}, makes the term $\sqrt{\mu}$
disappear from the formulation, so repulsive forces can be taken into
account.

The derivations of \citet{Pitkin_1965}\footnote{\citet{Pitkin_1965}
  calls universal variables the functions $U_n=\chi^n c_n$, which will
  be defined in Eq.~(\ref{eq:Un}) and named \textit{universal
    functions}.}, \citet{Sconzo_1967}, and \citet{EverhartPitkin_1983}
put in result the regularising role of the universal anomaly. By
applying the time transformation~(\ref{eq:chi}) and taking advantage
of the conservation of the energy, a third-order linear differential
equation with constant coefficients can be obtained for the position
vector ${\bf r}$ \citep[see][sect.~4.5]{Battin_1999}. In particular,
following more closely Stumpff's original approach,
\citet{Sconzo_1967} showed that Stumpff's functions can be introduced
in a straightforward way if the solution is represented via Taylor
series.

The universal elements ${\bf r}_0$, $\dot{{\bf r}}_0$ can
also describe rectilinear orbits, but the formulae that relate them to
${\bf r}$, $\dot{{\bf r}}$ lose their meaning when the
radial distance becomes zero, that is, when the body collides with the
centre of attraction.  In order to mitigate the loss of accuracy that
occurs close to the singularity, \citet{Pitkin_1965} suggested
replacing ${\bf r}_0$, $\dot{{\bf r}}_0$ by
${\bf r}_0/\sqrt{r_0}$, $r_0\dot{{\bf r}}_0/\sqrt{\mu}$,
respectively.

The variation of parameters equations for the universal variables
${\bf r}_0$, $\dot{{\bf r}}_0$ were first given by
\citet{Wong_1962} and later by
\citet[][sect.~16J]{Herrick_1965,Herrick_1971} and \citet{Pitkin_1966}
in a more suitable form for computer programming. The latter presented
some numerical tests with low-thrust trajectories.

This special perturbation method has two drawbacks from the
perspective of numerical integration. First, the variable $\chi$ must
be obtained at each step by solving the universal Kepler equation with
an iterative method. Moreover, the time derivatives of
${\bf r}_0$, $\dot{{\bf r}}_0$ contain secular terms in the
variable $\chi$. To solve this second problem, \citet{Born_etal_1974}
allowed $t_0$ to vary in a prescribed way instead of keeping it constant
throughout the propagation as in Herrick's variation of parameters
method.  However, one additional differential equation is required to
compute $\chi$, and the time derivatives of both $t_0$ and $\chi$ are
affected by secular terms. In fact, their elimination is not possible
without compromising the universality of the formulation
\citep[see][sect.~10.7]{Battin_1999}.\footnote{The secular terms are
  completely removed if, in addition to properly prescribing the
  variation of $t_0$, we include in the state vector the difference
  between the true anomalies of the current position at time $t$ and
  of the departure point at time $t_0$. The drawback of this approach
  is that the formulation becomes singular when the angular momentum
  vanishes, and therefore, it is not universal.}

There are other notable sets of elements that are universal, i.e.
well defined for any motion with the only possible exception of the
case $r=0$. They are related to the regularisations due to
\citet{Sperling_1961}, \citet[][hereafter KS]{KS_1965}, and to the
linearisation method shown by \citet{Burdet_1969}.

The \textit{natural} elements were derived by \citet{Burdet_1968} from
Sperling's regularisation. \citet{Sperling_1961} found that it is
possible to write a second-order linear differential equation not only
for the orbital radius $r$ but also for the position vector ${\bf r}$
if the eccentricity vector and the energy integral are both embedded
in the equation of motion resulting from the change of independent
variable~(\ref{eq:chi}). The solutions $r(\chi)$, ${\bf r}(\chi)$,
$t(\chi)$ are then expressed in terms of the natural elements and the
special functions $c_n$ (Eq.~\ref{eq:cn}) originally introduced by
Stumpff. The new formulation is universal, and it is valid even for
$r=0$. Burdet derived also the formulae for computing the variation of
the natural elements with respect to the anomaly $\chi$.
 
The general solution of the KS regularised equations in terms of
Stumpff's functions was presented by \citet{Deprit_1968}. The elements
that appear in the solution are uniformly valid for all values of the
Keplerian energy and are regular at collision. The same elements had
already been introduced by \citet{Broucke_1966}, who also obtained
explicit expressions of their derivatives by the method of variation
of parameters. \citet{Scheifele_1970},
\citet[][pp.~250--254]{StiSche_Book} applied the theory of
Hamilton--Jacobi to the KS Hamiltonian system to obtain two sets of
ten canonical elements that are regular and uniform with respect to
the total energy. An element linked to the physical time was naturally
introduced following this approach. \citet{Bond_1974} developed a
special perturbation method that is based on a set of elements very
similar to the one called Type II in \citet{Scheifele_1970}. An
alternative formulation in which mixed-secular terms are eliminated
from the derivative of the time element was also presented.

The idea behind the transformation applied by \citet{Burdet_1969}
dates back to the eighteenth century
\citep[see][]{Deprit_etal_1994}. The inverse of the orbital distance
($\rho$) and the radial unit vector (${\bf e}_r=\rho\,{\bf r}$) are
chosen as new coordinates to represent the position. Then, the system
of differential equations for $\rho$, ${\bf e}_r$ is linear if the
independent variable is changed according to the relation
\begin{equation}
  \kappa \mathrm{d}t=r^2\mathrm{d}\phi,\quad \kappa>0,
  \label{eq:phi}
\end{equation}
where $\kappa$ is a constant (at least of the Kepler problem). Burdet
chose $\kappa=1$, so that the frequency of oscillation of both $\rho$
and ${\bf e}_r$ along Keplerian motion is given by the angular
momentum of the particle divided by its mass ($h$).\footnote{This
  result for $\rho$ with $\kappa=h$ is called Binet's formula, after
  Jacques Binet (1786--1856), and it was already known to Isaac Newton
  (1642--1726).} The solution of the new system is written in a
unified way for $h>0$ and $h=0$ by means of functions that are
analogous to those used by Stumpff and of the \textit{focal}
elements. Their differential equations are derived together with that
of a time element.


\citet{Chelnokov_I} formally established the connection between KS
variables and the Euler parameters, already pointed out by
\citet{BrouckeLass1975}. These quantities represent a reference frame
that has one axis aligned with the position vector and rotates with
angular velocity always parallel to the angular momentum vector. By
changing time according to Eq.~(\ref{eq:phi}), the four Euler
parameters satisfy the equations of an harmonic oscillator with
frequency $1/2$ for $\kappa=h$. This fact opened the way for
generating new orbital elements, as shown by \citet{Chelnokov_II} and
more recently by \citet{RoaKasdin_2017}. We observe that the elements
proposed by these authors are universal if $\kappa=1$, but they are
not regular.

Although against the spirit of universal variables, we consider
formulations based on orbital elements that allow a uniform transition
through elliptic, parabolic, and hyperbolic motion as long as the
angular momentum is not zero. Milankovi{\'c}'s vectorial elements
describe the geometry of any orbit, and their definition is not related
to a particular reference frame
\citep{Milankovic_1939,AllanWard_1963}. Orbit propagation with these
quantities is possible thanks to appending to the state vector an
angle that locates the position of the particle with respect to a
preferably non-singular direction on the osculating plane. The true
longitude is suitable for this purpose
\citep{RoyMoran_1973,RosScheeres_2014}, but it loses its meaning when
$h=0$. Parameters related to an orbital reference frame\footnote{With
  the adjective \textit{orbital}, we mean that the reference frame is
  defined by the osculating plane of motion, and more specifically
  that one axis has the same direction of the angular momentum
  vector.} at epoch are doomed to fail in describing rectilinear
orbits. This is evident from the expression of the transverse unit
vector,
${\bf e}_{t,0}=[r_0\dot{{\bf r}}_0-({\bf r}_0\cdot\dot{{\bf r}}_0){\bf e}_{r,0}]/h$,
where ${\bf e}_{r,0}={\bf r}_0/r_0$. As noted by
\citet{Herrick_1965}, a proper scaling of ${\bf e}_{t,0}$ can
avoid the problem, and for example, ${\bf e}_{r,0}$,
$h{\bf e}_{t,0}$ recover their universal nature. The modified
equinoctial elements \citep{Walker_etal_1985} also fail at $h=0$
because they are related to an orbital reference frame
\citep{BrouckeCefola_1972}.

By setting $\kappa$ equal to $h$ in Burdet's
linearisation,\footnote{This method is known in the literature as
  Burdet--Ferr\'andiz regularisation. \citet{Ferrandiz_1988} achieved
  the same linearisation in the framework of the Hamiltonian formalism
  \citep[see][]{Deprit_etal_1994}.} the angle $\phi$ becomes the true
anomaly in the unperturbed case, and the oscillation frequencies of
$\rho$, ${\bf e}_r$ are equal to 1. This fact brings a considerable
advantage: the new elements do not exhibit secular terms in their
derivatives unlike the original focal elements. Even more interesting
is that three of them fix the shape of the osculating ellipse and the
remaining six define the orientation of an orbital reference frame,
which is called \textit{ideal} after \citet[][see p.~66 for the
  definition of the ideal coordinates]{Hansen_1857}. In the light of this
geometric interpretation, a reduction of the dimension of the system
from nine to seven is achieved by taking the Euler parameters that
describe the rotation of the ideal frame. The choice $\kappa=h$
introduces a singularity when the value of $h$ is zero which was not
present in Burdet's variables. \citet{Deprit_1975} and
\citet{Vitins_1978} developed seven elements of this kind by following
two different approaches. A review of several references about this
subject can be found in the introduction of \citet{Bau_etal_2015},
where the concept of Hansen ideal frames and the connection between
the ideal elements and Burdet's linearisation are discussed in
detail. The method named \textit{Dromo} \citep{Pelaez_etal_2007}
revived the interest in ideal elements for orbit propagation,
especially because the authors showed that it can be much more
accurate and faster than Cowell's method
\citep[][p.~447]{Battin_1999}. Dromo is based on seven quantities
almost equivalent to Deprit's and Vitins' and on a fictitious time
which is represented by the anomaly $\phi$ (Eq.~\ref{eq:phi}, wherein
$\kappa=h$). For an extensive presentation of Dromo, we refer to
\citet{Urrutxua_etal_2016} and \citet[][chap.~4]{Roa_Book}, who also
mention the important updates that have been recently proposed to
improve its numerical performance.

A propagator similar to Dromo but working only for negative values of
the total energy was devised by \citet{Bau_etal_2014,
  Bau_etal_2015}. The basic idea behind this method is to search for a
linearisation of the equations of motion starting from the
\textit{projective} coordinates $(r,{\bf e}_r)$, as in the
Burdet--Ferr\'andiz regularisation, and choosing a time transformation
of Sundman's type instead of Eq.~(\ref{eq:phi}). In the unperturbed
motion, the independent variable is the eccentric anomaly and the
differential equation of the radial distance $r$ is linear with
constant coefficients \citep[a well-known result,
  see][]{Bohlin_1911}. The solution can be written so that the two
constants of integration are the projections of the eccentricity
vector along a pair of fixed orthogonal axes which lie on the orbital
plane. Based on these two directions, the authors defined a reference
frame, named \textit{intermediate}, and introduced four Euler
parameters to represent its orientation in space. The six integrals of
the Kepler problem obtained in this way constitute the state vector
together with the semi-major axis, and a time element. The special
perturbation method generated from the new elements can exhibit a
substantial advantage with respect to Dromo. Numerical investigations
conducted by \citet{Amato_etal_2017,Amato_etal_2019} show its
excellent behaviour in the propagation of both asteroids and
artificial satellites of the Earth. An analogous formulation was
derived independently by \citet{RoaPelaez_2015} and
\citet{Bau_etal_2016} for positive values of the Keplerian and total
energy, respectively.

The methods proposed by \citet{Bau_etal_2015, Bau_etal_2016},
\citet{RoaPelaez_2015} cannot be used, in general, to propagate a body
that presents transitions from elliptic to hyperbolic motion or vice
versa. We also expect that they loose accuracy when the energy is
close to zero. Therefore, we tried to find a unique formulation that
includes those in \citet{Bau_etal_2015, Bau_etal_2016} and
\citet{RoaPelaez_2015} and is able to deal with cases in which the
Keplerian or the total energy changes sign during the motion. In the
present paper we achieve such goal by switching to a regularising time
variable and taking advantage of Stumpff's functions as shown in
Section~\ref{sec:ProEl}. Eight uniform elements arise from our
procedure: they are non-singular for any value of the total energy and
are not defined when the angular momentum is zero. The new quantities
are called \textit{intermediate elements} because there exists an
intermediate frame that plays a key role in their definition. This
frame establishes the orientation of the osculating plane and of a
departure direction on it, from which the position of the particle is
reckoned. In Section~\ref{sec:ProEl}, we also obtain the first-order
differential equations that govern the variations of the new elements
with respect to the fictitious time. Numerical tests assessing the
performance of the new method for orbit propagation are shown in
Section~\ref{sec:num_res}.

In our derivation, particular attention is paid to the appearance of
secular terms with respect to the independent variable when the total
energy is negative. In order to better understand their origin, we
introduce an arbitrary quantity $\beta$ in the time transformation
(Eq.~\ref{eq:fict_time}). Secular terms can only be eliminated from
the time derivatives of the new elements by selecting $\beta$ in a
proper way, at the cost of losing the uniform character of the
proposed special perturbation method. An alternative formulation which
is completely free of secular terms and that is still uniform is
derived in Appendix~\ref{sec:nosec} after Conclusions (in
Section~\ref{sec:thend}).

In some applications, orbit propagation is part of a more complicated
procedure known as orbit determination: given a set of observations at
different epochs relative to the same celestial body, we want to
determine its position and velocity and the associated uncertainties
at some prescribed epoch. An essential ingredient in orbit
determination is the \textit{state-transition matrix} (STM). Its
elements are the partial derivatives of position and velocity with
respect to their initial values and obey the \textit{variational
  equation}. \citet{Sitarski_1967} presented a solution of the
two-body variational equation which is independent of the type of the
orbit. \citet{Crawford_1969} started from Sitarski's result to write a
simpler expression of the two-body STM.  \citet{Herrick_1965} and
\citet{Goodyear_1965, Goodyear_1966} derived a closed-form solution
for the partial derivatives in terms of Stumpff's functions \citep[a
  simpler presentation than Herrick's is available
  in][sect.~9.7]{Battin_1999}. Improvements to Goodyear's STM were
proposed by \citet{Shepperd_1985} and \citet{Der_1997}: the former
suggested a new scheme for solving Kepler's equation, which is a
preliminary step necessary to compute the STM; the latter found a way
to remove the secular terms contained in the universal functions $U_4$
and $U_5$ (see Eq.~\ref{eq:Un}).

Section~\ref{sec:DC} deals with the use of our formulation for orbit
determination and uncertainty propagation.  The delicate aspect of
computing the STM for the rectangular coordinates at a certain time
from the STM of the intermediate elements at the corresponding
fictitious time is addressed. Appendix~\ref{sec:dfdi} reports the
expressions of the derivatives that appear in the variational
equations of the intermediate elements.

\section{The intermediate elements}
\label{sec:ProEl}
Consider the perturbed Kepler problem. The evolution of the position
${\bf r}$ of a particle of mass $m$ with respect to a body of mass $M$
(here referred to as the central body) is described by Newton's
second law
\begin{equation}
  \ddot{{\bf r}}=-\frac{\mu}{r^{3}}{\bf r}+{\bf F},
  \label{eq:2bp}
\end{equation}
where $\mu=G(m+M)$, with $G$ the gravitational constant, $r=|{\bf r}|$
and ${\bf F}$ is the vector sum of the perturbing forces acting on
$m$. We assume that
\begin{equation}
  {\bf F}({\bf r},\dot{{\bf r}},t)={\bf P}({\bf r},\dot{{\bf r}},t)
  -\nabla\Up({\bf r},t),
  \label{eq:F}
\end{equation}
where $\nabla\Up$ is the gradient of the disturbing potential $\Up$
and ${\bf P}$ is the sum of the perturbing forces that are not related
to the gradient of a potential energy. We will refer to
\textit{unperturbed motion} when both ${\bf F}={\bf 0}$ and $\Up=0$.

For future use, let us introduce the local vertical, local horizontal
(LVLH) reference frame $\{O,{\bf e}_r,{\bf e}_{\nu},{\bf e}_z\}$,
where $O$ denotes the location of the centre of mass of the central
body, and
\begin{equation}
  {\bf e}_r=\frac{{\bf r}}{r},\qquad {\bf e}_{\nu}={\bf e}_z\times{\bf e}_{r},
  \qquad {\bf e}_z=\frac{{\bf r}\times\dot{{\bf r}}}{|{\bf r}\times\dot{{\bf r}}|}.
\label{eq:ernuz}
\end{equation}
In this section, we develop a set of eight orbital elements that can be
used to represent the position and velocity of the particle at a given
epoch. We first derive the elements that describe the motion
\textit{on} the orbital plane and next those that describe the
evolution \textit{of} the orbital plane.

\subsection{Motion {\textit{on}} the orbital plane}
\label{sec:Mot_onplane}
\begin{figure}
  \begin{center}
    \begin{tikzpicture}
      \coordinate (Or) at (3.5,0);
      \coordinate (O) at (0,0);
      \coordinate (Ou) at (0,3.5);
      \draw [-latex] (O) -- (Or);
      \draw [-latex] (O) -- (Ou);
      \coordinate (G) at ($({3*cos(45)},{3*sin(45)})$);
      \coordinate (R) at ($({5*cos(60)},{5*sin(60)})$);
      \draw [-latex] (O) -- (G);
      \draw [-latex] (O) -- (R);      
      \draw (1,0) arc (0:60:1);
      \coordinate (nu) at ($({1*cos(22.5)},{1*sin(22.5)})$);
      \draw ($({1.5*cos(45)},{1.5*sin(45)})$) arc (45:60:1.5);
      \coordinate (th) at ($({1.5*cos(52.5)},{1.5*sin(52.5)})$);
      \draw (O) node [left,below] {$O$};
      \draw (Or) node [right,above] {${\bf e}_x$};
      \draw (Ou) node [left,above] {${\bf e}_y$};
      \draw (G) node [right,above] {${\bf g}$};
      \draw (R) node [right,above] {${\bf r}$};
      \draw (th) node [right=1mm,above] {$\theta$};
      \draw (nu) node [right=1mm,above] {$\nu$};    
    \end{tikzpicture}
  \end{center}
  \caption{Orientation of the position vector ${\bf r}$ and the
    generalised eccentricity vector ${\bf g}$ with respect to the
    unit vectors ${\bf e}_x$, ${\bf e}_y$ of the
    intermediate frame. All vectors lie on the osculating orbital
    plane at some epoch $t$.}
  \label{fig:1}
\end{figure}
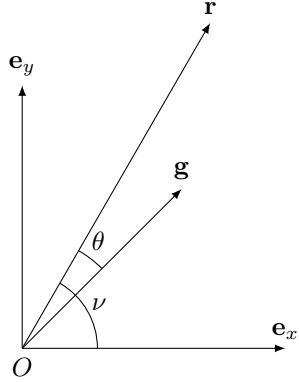

Let us introduce the polar coordinates $(r,\nu)$ on the osculating
plane of motion, where $\nu$ is the angle measured from a reference
axis $Ox$ to the position vector ${\bf r}$. The definition of
$\nu$ and therefore of $Ox$ is given in the end of this section.
Then, we can introduce the {\it intermediate} reference frame
$\{O,{\bf e}_x,{\bf e}_y,{\bf e}_z\}$, where
${\bf e}_x$ obeys the relations
${\bf e}_x\cdot{\bf e}_r=\cos\nu$,
${\bf e}_x\times{\bf e}_r=\sin\nu\,{\bf e}_z$ (see
Figure~\ref{fig:1}), and
${\bf e}_y={\bf e}_z\times{\bf e}_x$.

The independent variable is changed from the physical time $t$ to a
fictitious time $\chi$ by the transformation
\begin{equation}
\beta \mathrm{d}t=r\,\mathrm{d}\chi,\label{eq:fict_time}
\end{equation}
where $\beta\in\mathbb{R}^{+}$ is an arbitrary constant along any
solution of the Kepler problem. The introduction of the parameter
$\beta$ has been suggested in the past, for example, by
\citet{Herrick_1965}. We will make a specific choice for $\beta$ in
Section~\ref{sec:Prop_form}. The orbital radius obeys the second-order
differential equation
\begin{equation}
\beta^{2}r''=2\TE r+\mu+r(rF_{r}-2\Up)-\beta
r'\beta',\label{eq:d2r_dchi2}
\end{equation}
where prime denotes differentiation with respect to $\chi$, $F_{r}$ is
the radial component of the perturbing force ${\bf F}$, and $\TE$ is
the specific total energy. Equation~(\ref{eq:d2r_dchi2}) is obtained
from~(\ref{eq:2bp}) and~(\ref{eq:fict_time}). The quantity $\TE$ is
defined as
\begin{equation}
  \TE=\frac{1}{2}\Bigl(\dot{r}^{2}+\frac{h^{2}}{r^{2}}\Bigr)
  -\frac{\mu}{r}+{\Up},\label{eq:tot_en}
\end{equation}
where $h=|{\bf r}\times\dot{{\bf r}}|$ is the specific angular
momentum. From~(\ref{eq:d2r_dchi2}), one finds that the following
relation holds for the Kepler problem
\begin{equation}
  \sigma''=-\alpha\sigma,\label{eq:d2sigma}
\end{equation}
where $\sigma=r'$, and
\begin{equation}
  \alpha=-\frac{2\TE}{\beta^{2}}.
\end{equation}

Let us introduce the {\it universal functions}
\citep[][sect.~4.5]{Battin_1999}:
\begin{equation}
  U_n(\chi;\alpha)=\chi^nc_n(\chi;\alpha),\quad n\in\mathbb{N},
  \label{eq:Un}
\end{equation}
where
\begin{equation}
  c_n(\chi;\alpha)=\sum_{k=0}^{\infty}(-1)^{k}
  \frac{(\alpha\chi^2)^k}{(n+2k)!}.
  \label{eq:cn}
\end{equation}
The series $c_n(\chi;\alpha)$, known as Stumpff's functions, converge
absolutely for all values of $\chi$, $\alpha$, and uniformly in any
bounded domain of $\chi$, $\alpha$. The solution of
Eq.~(\ref{eq:d2sigma}) can be written as
\begin{equation}
  \sigma=a_1U_0(\chi;\alpha)+a_2U_1(\chi;\alpha).\label{eq:sigma1}
\end{equation}
Then, the orbital radius adopts the form
\begin{equation}
  r=a_0+a_1U_1(\chi;\alpha)+a_2U_2(\chi;\alpha).\label{eq:r1}
\end{equation}
The constants of integration $a_0$, $a_1$, $a_2$ are determined from
the initial values of $r$, $\sigma$. If we assume that $\chi=0$ at the
initial time, we obtain
\begin{equation}
  a_0=r(0)=r_{0},\qquad a_1=\sigma(0)=\sigma_0.\label{eq:a0a1}
\end{equation}
By evaluating the equation $\beta^{2}r''=2\TE r+\mu$ at
$\chi=0$, we also find
\begin{equation}
  a_2=\frac{\mu+2\TE r_0}{\beta^{2}}.\label{eq:a2}
\end{equation}
Using~(\ref{eq:a0a1}),~(\ref{eq:a2})
in~(\ref{eq:sigma1}),~(\ref{eq:r1}), and noting that $U_0+\alpha
U_2=1$, we can write
\begin{align}
  r & = r_0U_0(\chi;\alpha)+\sigma_0U_1(\chi;\alpha)+\frac{\mu}{\beta^2}U_2(\chi;\alpha),
  \label{eq:r2}\\[3pt]
  \sigma & = \sigma_0U_0(\chi;\alpha)+\frac{\mu}{\beta^{2}}(1-\lambda r_0)U_1(\chi;\alpha),
  \label{eq:sigma2}
\end{align}
where
\begin{equation}
  \lambda=\frac{-2\TE}{\mu}.
\end{equation}
The quantities $r_0$, $\sigma_0$ will not be constant if perturbations
are present (${\bf F}\ne{\bf 0}$). Moreover, the time evolution
of $r_0$, $\sigma_0$ will depend on the specific choice of $\beta$,
that is on the choice of independent variable (see
Eq.~\ref{eq:fict_time}).

Before dealing with the polar angle $\nu$, we define
\begin{equation}
  g=\sqrt{1-\lambda p},\qquad \mu p=h^{2}+2r^{2}{\Up},\qquad
  c=\sqrt{\mu p}.\label{eq:gpc}
\end{equation}
Let us call $g$, $p$, $c$ the \textit{generalised} eccentricity,
semilatus rectum, and angular momentum, respectively. They reduce to
their \textit{osculating} counterparts when $\Up=0$. The quantities
$g$, $p$ are functions of $r_0$, $\sigma_0$, $\beta$, $\lambda$. Their
expressions, which can be found from~(\ref{eq:gpc})
and~(\ref{eq:tot_en}),~(\ref{eq:r2}),~(\ref{eq:sigma2}), are
\begin{align}
  g^2 & = (1-\lambda r_0)^2+\frac{\beta^2}{\mu}\lambda\sigma_0^2,\label{eq:g2}\\[3pt]
  p & =r_0(2-\lambda r_0)-\frac{\beta^2}{\mu}\sigma_0^2.\label{eq:p}
\end{align}
Note that $g$, $p$ take finite values for $\alpha=0$.

We aim at relating the polar angle $\nu$ to the independent variable
$\chi$. The angle $\nu$ is comprised between the position of the
particle and a reference axis that lies on the osculating orbital
plane and passes through the central body. Additionally, this axis
must remain fixed in space at least when the motion is unperturbed and
to be well defined for any value of $h$ different from zero. Possible
definitions of the polar angle $\nu$ must obey the following
condition: when the motion is Keplerian, $\nu$ is the true anomaly up
to an additive constant angle, that is
\begin{equation}
  {\dot\nu}=\frac{c}{r^2}.\label{eq:dnudt}
\end{equation}
In our formulation, $\nu$ is defined as follows. Assume $\nu(0)=0$,
then from Eqs.~(\ref{eq:dnudt}) and~(\ref{eq:fict_time}) we have
\begin{equation}
  \nu=\frac{c}{\beta}\int_0^\chi \frac{1}{r(s)} \mathrm{d}s,
  \label{eq:nu_int}
\end{equation}
with $r(s)$ taken from Eq.~(\ref{eq:r2}). After solving the
integral, we find
\begin{equation}
  \beta\tan\frac{\nu}{2}=\frac{c\,U_1\left(\frac{1}{2}\chi;\alpha\right)}{r_0
    U_0\left(\frac{1}{2}\chi;\alpha\right)+\sigma_0 U_1\left(\frac{1}{2}\chi;\alpha\right)}.
  \label{eq:Gauss_eq}
\end{equation}
The above formula, called by \citet{Sperling_1961} the Gaussian
equation, defines $\nu$ as a function of $\chi$.

\subsection{Particularisations for positive, negative, and zero values of $\alpha$}
\label{sec:cases_on_plane}
Equations~(\ref{eq:r2}),~(\ref{eq:Gauss_eq}) are here particularised
to the cases $\alpha>0$, $\alpha<0$, $\alpha=0$, which correspond to
negative, positive, and zero total energy,
respectively.
For this purpose, we need to provide the definition of the generalised
true anomaly $\theta$. Given the generalised eccentricity vector 
\begin{equation}
  {\bf g}={\bf w}\times({\bf r}\times{\bf w})-{\bf e}_r,\label{eq:g}
\end{equation}
where
\begin{equation}
  {\bf w}=\dot{r}\,{\bf e}_r+\frac{c}{r}\,{\bf e}_{\nu},
\end{equation}
we have that $\theta$ is the angle measured from ${\bf g}$ to
${\bf r}$ counterclockwise as seen from ${\bf e}_z$ (see
Figure~\ref{fig:1}). Therefore, from~(\ref{eq:g}) and noting that
$|{\bf g}|=g$, with $g$ given in~(\ref{eq:gpc}), we have ($c\ne
0$)
\begin{equation}
  g\cos\theta=\frac{c^2}{r}-1,\qquad g\sin\theta=c\,\dot{r}.\label{eq:theta}
\end{equation}

\subsubsection{The case $\alpha>0$}
After substituting into Eq.~(\ref{eq:r2}) the expressions taken
by $U_{0}$, $U_{1}$, $U_{2}$ for $\alpha>0$ we have
\begin{equation}
  r=\frac{1}{\lambda}[1-\wp_{1}\cos(\sqrt{\alpha}\chi)-\wp_{2}\sin(\sqrt{\alpha}\chi)],
  \label{eq:r_ell}
\end{equation}
where
\begin{equation}
  \wp_1=1-\lambda r_0,\qquad \wp_2=-\beta\sigma_0\sqrt{\frac{\lambda}{\mu}}.
\end{equation}
Then, following \citet{Bau_etal_2015} we can define the
generalised eccentric anomaly $G$ from
\begin{equation}
  \mu g\cos G=\mu+2\TE r,\qquad \mu g\sin G=r\,\dot{r}\sqrt{-2\TE}.\label{eq:G}
\end{equation}
By using the first relation in~(\ref{eq:G}) and Eq.~(\ref{eq:r_ell}),
we obtain
\begin{equation}
  \begin{split}
    \wp_1 & = g\cos(\sqrt{\alpha}\chi-G),\\[3pt]
    \wp_2 & = g\sin(\sqrt{\alpha}\chi-G).
  \end{split}
\end{equation}
Since $\wp_1=g\cos G_0$, $\wp_2=-g\sin G_0$, where $G_0$ is the value
taken by $G$ at $\chi=0$, we have
\begin{equation}
  \sqrt{\alpha}\chi=G-G_0.
\end{equation}  
Equation~(\ref{eq:Gauss_eq}) can be written for
$\alpha>0$ as
\begin{equation}
  \tan\frac{\nu}{2}=\frac{\sqrt{1-\wp_1^2-\wp_2^2}\sin\left(\frac{1}{2}\sqrt{\alpha}\chi\right)}
  {(1-\wp_1)\cos\left(\frac{1}{2}\sqrt{\alpha}\chi\right)-\wp_2\sin\left(\frac{1}{2}\sqrt{\alpha}\chi\right)},
  \label{eq:GE_ell}
\end{equation}
or alternatively as
\begin{equation}
  \tan{\frac{\nu-\theta}{2}}=\sqrt{\frac{1+g}{1-g}}\tan\frac{\sqrt{\alpha}\chi-G}{2}.
  \label{eq:GE_ell1}
\end{equation}
Thus, the angular difference $\nu-\theta$ is obtained from
$\sqrt{\alpha}\chi-G$ by applying the classical relation between the
true anomaly and the eccentric anomaly in the two-body problem.

\noindent\textit{Remark.} The method presented in \citet{Bau_etal_2015},
called EDromo, employs $\beta=\sqrt{-2\varepsilon}$, so that
$\alpha=1$ and Eq.~(\ref{eq:r_ell}) becomes
\begin{equation*}
  r=\frac{1}{\lambda}(1-\wp_1\cos\chi-\wp_2\sin\chi),
\end{equation*}
where\footnote{In \citet{Bau_etal_2015} the elements $\wp_1$, $\wp_2$,
  $\lambda^{-1}$ are denoted by $\lambda_1$, $\lambda_2$, $\lambda_3$,
  respectively, and the independent variable $\chi$ by $\varphi$.}
\begin{equation}
  \wp_1=g\cos(\chi-G),\qquad \wp_2=g\sin(\chi-G).\label{eq:wp12_ell}
\end{equation}
Interestingly, in EDromo the polar angle is defined as
\begin{equation}
  \nu=\theta+\chi-G.\label{eq:GE_ell2}
\end{equation}
This choice seems quite natural looking at the expressions of $\wp_1$,
$\wp_2$ given in~(\ref{eq:wp12_ell}), which suggest to directly take
$\chi-G$ as the angle between ${\bf g}$ and ${\bf e}_x$ (see
Figure~\ref{fig:1}). Finally, it is worth noting that
from~(\ref{eq:GE_ell2}) by using a formula proposed by
\citet{BrouckeCefola_1973}, Eqs.~(\ref{eq:G}) and the first relation
in~(\ref{eq:gpc}), we get the relation
\begin{equation*}
  \nu=\chi+2\arctan\frac{\sigma}{r+\sqrt{p/\lambda}}.
\end{equation*}

\subsubsection{The case $\alpha<0$}
For $\alpha<0$, Eq.~(\ref{eq:r2}) becomes
\begin{equation}
  r=-\frac{1}{\lambda}[\wp_1\cosh(\sqrt{-\alpha}\chi)+\wp_2\sinh(\sqrt{-\alpha}\chi)-1],
  \label{eq:r_hyp}
\end{equation}
where
\begin{equation}
  \wp_{1}=1-\lambda r_0,\qquad \wp_{2}=\beta\sigma_0\sqrt{\frac{-\lambda}{\mu}}. 
\end{equation}
We can define the generalised hyperbolic anomaly $F$ by 
\citep[see][]{Bau_etal_2016}
\begin{equation}
  \mu g\cosh F=\mu+2\TE r,\qquad \mu g\sinh F=r\,\dot{r}\sqrt{2\TE},\label{eq:gch_gsh}
\end{equation}
where $g$ is the generalised eccentricity as in~(\ref{eq:G}). Then,
the following relations hold:
\begin{equation}
  \begin{split}
    & \wp_{1}=g\cosh(F-\sqrt{-\alpha}\chi),\\[3pt]
    & \wp_{2}=g\sinh(F-\sqrt{-\alpha}\chi).
  \end{split}
\end{equation}
Since $\wp_1=g\cosh F_0$, $\wp_2=g\sinh F_0$, where $F_0$ is the value
of $F$ for $\chi=0$, we have
\begin{equation}
  \sqrt{-\alpha}\chi=F-F_0.
\end{equation}
Equation~(\ref{eq:Gauss_eq}) for $\alpha<0$ can be written as
\begin{equation}
  \tan\frac{\nu}{2}=\frac{\sqrt{\wp_1^2-\wp_2^2-1}\sinh\left(\frac{1}{2}\sqrt{-\alpha}\chi\right)}
  {(\wp_1-1)\cosh\left(\frac{1}{2}\sqrt{-\alpha}\chi\right)+\wp_2\sinh\left(\frac{1}{2}\sqrt{-\alpha}\chi\right)},
  \label{eq:GE_hyp}
\end{equation}
or alternatively as
\begin{equation}
  \tan{\frac{\nu-\theta}{2}}=\sqrt{\frac{1+g}{g-1}}\tanh\frac{\sqrt{-\alpha}\chi-F}{2},
  \label{eq:GE_hyp1}
\end{equation}
which is the relation between the true anomaly and the hyperbolic
anomaly in the two-body problem.

\noindent\textit{Remark.} The method presented in
\citet{Bau_etal_2016}, here called HDromo, employs
$\beta=\sqrt{2\varepsilon}$, so that $\alpha=-1$ and
Eq.~(\ref{eq:r_hyp}) takes the form
\begin{equation*}
  r=-\frac{1}{\lambda}(\wp_{1}\cosh\chi+\wp_{2}\sinh\chi-1),
\end{equation*}
where\footnote{In \citet{Bau_etal_2016} the elements $\wp_1$, $\wp_2$,
  $-\lambda^{-1}$ are denoted by $\lambda_1$, $\lambda_2$, $\lambda_3$,
  respectively, and the independent variable $\chi$ by $\varphi$.}
\begin{equation*}
  \wp_{1}=g\cosh(F-\chi),\qquad \wp_{2}=g\sinh(F-\chi).
\end{equation*}
HDromo implements a different definition for $\nu$ with respect to
that given in Eqs.~(\ref{eq:GE_hyp}),~(\ref{eq:GE_hyp1}). Let us
consider
\begin{equation*}
  \zeta={\rm gd}\,2(F-\chi),\qquad \tan\frac{\zeta}{2}=\tanh(F-\chi),
\end{equation*}
where ${\rm gd}\,x$ is the Gudermannian function
\citep[see][p.165]{Battin_1999}. Then, we obtain for $\wp_1$, $\wp_2$:
\begin{equation*}
  \wp_{1}=\wp\cos\frac{\zeta}{2},\qquad \wp_{2}=\wp\sin\frac{\zeta}{2},
\end{equation*}
where $\wp=\sqrt{{\wp_1}^2+{\wp_2}^2}$. These expressions invite to
set the angle between ${\bf g}$ and the reference axis equal to
${\zeta}/{2}$, so that
\begin{equation*}
  \nu=\theta+\frac{\zeta}{2}.
\end{equation*}

\subsubsection{The case $\alpha=0$}
For $\alpha=0$, the equations for the orbital radius and the polar
angle reduce to
\begin{equation}
  r=r_0+\sigma_0\chi+\frac{\mu}{2\beta^2}\chi^{2},\qquad
  \beta\tan\frac{\nu}{2}=\frac{c\,\chi}{2r_0+\sigma_0\chi}.
\end{equation}
One can find that $\chi$, $\nu$ are related to the angle $\theta$,
introduced in~(\ref{eq:theta}). Indeed,
\begin{align}
  \chi & = \frac{c\,\beta}{\mu}\left(\tan\frac{\theta}{2}-\tan\frac{\theta_0}{2}\right),\\[3pt]
  \nu & = \theta-\theta_0.
\end{align}
In fact, the latter relation holds for any $\alpha$.

\subsection{Variation of the elements $r_0$, $\sigma_0$, $\lambda$}
\label{sec:VOP_rsl0}
The quantities $r_0$, $\sigma_0$, $\lambda$ are attractive candidates
for the set of intermediate elements. From their evolution, one
obtains $r$, $\sigma$, $p$, $\nu$ as functions of $\chi$ once $\beta$
is defined. Then, from the orientation of the intermediate basis
$\{{\bf e}_x,{\bf e}_y,{\bf e}_z\}$, we can determine
the position and velocity of the particle at any $\chi$. In this
section, we deal with the computation of the derivatives of $r_0$,
$\sigma_0$, $\lambda$ with respect to $\chi$, which vanish when the
motion is unperturbed. For simplicity, we will adopt the notation
\begin{equation}
  U_n=U_n(\chi;\alpha),\qquad \tilde{U}_n=U_n(2\chi;\alpha),\qquad
  n\in\mathbb{N}.
\end{equation}

Consider the equations
\begin{align}
  \frac{\mathrm{d}r}{\mathrm{d}\chi} & = \sigma,\\[3pt]
  \frac{\mathrm{d}\sigma}{\mathrm{d}\chi} & = \frac{1}{\beta^2}[2\TE
  r+\mu+r(rF_{r}-2\Up)]-\frac{\sigma\beta'}{\beta},
\end{align}
which stem from differentiating the definition of $r$, $\sigma$ given
in~(\ref{eq:r2}),~(\ref{eq:sigma2}), and regarding $r_0$, $\sigma_0$,
$\lambda$, $\beta$ as functions of $\chi$. After some algebraic
manipulations, we find
\begin{align}
  \beta^2r_0' & = -r(rF_{r}-2\Up)U_1-\frac{\mu}{4}\Big(r_0\tilde{U}_2+\sigma_0\tilde{U}_3
  +2\frac{\mu}{\beta^2}U_2^2\Big)\lambda'\nonumber\\
  &\quad\, +\chi\sigma_0\beta\beta',
  \label{eq:r0p}\\[3pt]
  \beta^2\sigma_0' & =
  r(rF_{r}-2\Up)U_0+\frac{\mu}{4}\Big[r_0(2\chi+\tilde{U}_1)+\sigma_0\tilde{U}_2\nonumber\\
  &\quad\, +\frac{\mu}{\beta^2}(\tilde{U}_3-4U_3)\Big]\lambda'
  +\Big[\frac{\mu}{\beta^2}(1-\lambda r_0)\chi-\sigma_0\Big]\beta\beta'.
  \label{eq:sigma0p}
\end{align}
Moreover, we have
\begin{equation}
  \lambda'=-\frac{2}{\mu}\Big(\sigma P_r +\frac{h}{\beta}P_{\nu}+
  \frac{r}{\beta}\frac{\partial\Up}{\partial t}\Big),
\end{equation}
where $P_r={\bf P}\cdot{\bf e}_r$,
$P_{\nu}={\bf P}\cdot{\bf e}_{\nu}$.

It is worth noting that for $\TE<0$, the expressions of $r_0'$,
$\sigma_0'$ contain some terms in which $\chi$ appears explicitly. The
presence of these terms can deteriorate the accuracy of $r_0$,
$\sigma_0$ computed by numerical integration of
Eqs.~(\ref{eq:r0p}),~(\ref{eq:sigma0p}), especially for long
propagations. On the other hand, the variational equations of $g$, $p$
(see~\ref{eq:g2},~\ref{eq:p}) are not affected by this
disadvantage. Secular terms can be avoided in both $r_0'$ and
$\sigma_0'$ if and only if we select $\beta=k\sqrt{-2\TE}$ ($\TE<0$),
where $k$ is a nonzero constant, so that the quantity $\alpha$ is
conserved along the perturbed motion.\footnote{In the method EDromo
  \citep{Bau_etal_2015}, it is $\alpha=1$ and secular terms are not
  present in the derivatives of $\wp_1$, $\wp_2$
  (see~\ref{eq:wp12_ell}).}

In Appendix~\ref{sec:nosec}, we show that it is still possible to
eliminate the secular terms from the derivatives $r_0'$, $\sigma_0'$
without having to restrict the domain of $\TE$ to negative values, by
adequately changing Eq.~(\ref{eq:d2r_dchi2}) and imposing that $\beta$
is constant, i.e. $\beta'=0$.

\subsection{The time element $t_0$ and its evolution}
\label{sec:Kep_eq}
By integrating the time transformation~(\ref{eq:fict_time}), we obtain
Kepler's equation in its universal form
\begin{equation}
  \beta(t-t_0)=r_0U_1(\chi;\alpha)+\sigma_0U_2(\chi;\alpha)
  +\frac{\mu}{\beta^2}U_3(\chi;\alpha),\label{eq:kep_eq}
\end{equation}
where the quantity $t_0$ is called {\it time element}.

In the classic formulations by \citet{Wong_1962},
\citet{Herrick_1965}, and \citet{Pitkin_1966}, time is the independent
variable and the value of $\chi$ corresponding to a given $t$ is
obtained by solving the universal Kepler equation.
Moreover, the time element $t_0$ is a constant, also when the motion
is perturbed.  \citet{Born_etal_1974} suggested that it may be more
convenient to let $t_0$ vary with time instead of keeping it fixed. In
fact, by properly choosing the time derivative of $t_0$, the secular
terms that appear in the variational equations of ${\bf r}_0$,
$\dot{{\bf r}}_0$ can be eliminated.\footnote{Unfortunately, the
  secular terms are not completely removed since they are contained in
  the expression of $\dot{t}_0$ \citep[see][pp.~510,
    511]{Battin_1999}.} The drawback of this approach is that $t_0$
is added to the state vector thus increasing the dimension of the
system.

In our formulation, $\chi$ is the independent variable as defined
in~(\ref{eq:fict_time}), and Eq.~(\ref{eq:kep_eq}) is used to directly
compute the physical time $t$, so $t_0$ must be known. The variational
equation of $t_0$, which is obtained by differentiation of
Eq.~(\ref{eq:kep_eq}), becomes
\begin{multline}
  \beta^3t'_0=r(rF_r-2\Up)U_2-\frac{\mu}{4}\Big[r_0(4U_3-\tilde{U}_3)-2\sigma_0U_2^2\\
  -\frac{\mu}{\beta^2}(\tilde{U}_5-8U_5)\Big]\lambda'+\chi r_0\beta\beta'.
\end{multline}
In the case $\TE<0$, the expression above for $t'_0$ contains terms
that are linear in $\chi$.\footnote{Note that the terms with
  $\chi^3$, which stem from $\tilde{U}_5$, $U_5$, cancel out.}  By
selecting $\beta=k\sqrt{-2\TE}$, where $k$ is a nonzero constant (see
Section~\ref{sec:VOP_rsl0}), we can eliminate only some of them,
because, as expected, those in $\tilde{U}_5-8U_5$ survive. However,
with this choice of $\beta$ it is still possible to get rid of the
secular terms as follows. Let us use the identity $U_1+\alpha U_3=\chi$
to write Kepler's equation as
\begin{equation}
  k\sqrt{\lambda\mu}(t-t_1)=\Big(r_0-\frac{1}{\lambda}\Big)U_1+\sigma_0U_2,
  \label{eq:kep_eq1}
\end{equation}
where
\begin{equation}
  t_1=t_0+\frac{\chi}{k\sqrt{\lambda^3\mu}}.
\end{equation}
Then, the variational equation of $t_1$ is free of secular terms. This
quantity, which is a linear function of $\chi$ when the motion is
unperturbed, is also referred to as {\it linear} time element. In
\citet{Bau_etal_2015}, both $t_0$ and $t_1$ are presented for EDromo.

A time element analogous to $t_0$ was developed also by
\citet{Burdet_1968} and \citet{Bond_1974}. In their formulation, the
time transformation~(\ref{eq:fict_time}) is applied with $\beta$ equal
to $\sqrt{\mu}$ and $1$, respectively. Moreover, the variables $r_0$,
$\sigma_0$ are included in the set of elements, even if they are not
necessary to describe the motion. The advantage of adding these
redundant variables is not clear in \citet{Burdet_1968}. On the other
hand, Bond finds out that substituting in~(\ref{eq:fict_time}) the
expression of $r$ given in~(\ref{eq:r2}), $t'_0$ is not affected by
{\it mixed} secular terms.\footnote{These terms contain the product
  of a trigonometric function of $\chi$ and some power of $\chi$.}
They arise instead if the orbital distance is written as a function of
the elements related to the Kustaanheimo-Stiefel parameters.

\subsection{Motion \textit{of} the orbital plane}
\label{sec:Mot_ofplane}
The proposed method relies on the existence of the orbital plane, and
so it becomes singular when the angular momentum vanishes. We track
the evolution of this plane and of a reference direction on it by
describing the orientation of the {\it intermediate} frame
$\{O,{\bf e}_x,{\bf e}_y,{\bf e}_z\}$ that we have
introduced in Section~\ref{sec:Mot_onplane}.

Let $\{O,{\bf e}_1,{\bf e}_2,{\bf e}_3\}$ be a
reference frame with the origin at $O$ (i.e. the centre of mass of
the central body) and the directions of ${\bf e}_i$,
$i=1,\,2,\,3$, fixed in space. In particular, the unit vectors
${\bf e}_1$, ${\bf e}_2$ generate the fundamental plane
(e.g. the plane of the Earth's orbit, or the plane of the Earth's
equator). We denote by $\Omega$, $I$, $\omega$ the three classical
orbital elements given by the longitude of the ascending node,
inclination, and argument of pericentre. Let us consider the quantity
$\Psi=\omega+f-\nu$, where $f$ is the true anomaly and $\nu$ is the
angle defined in Section~\ref{sec:Mot_onplane}. Then, the Euler angles
$\Omega$, $I$, $\Psi$ define the orientation of the basis
$\{{\bf e}_x,{\bf e}_y,{\bf e}_z\}$ with respect to
$\{{\bf e}_1,{\bf e}_2,{\bf e}_3\}$.

Following \citet[][p.~155]{Goldstein_1980}, we introduce the Euler
parameters $q_1$, $q_2$, $q_3$, $q_4$ related to the Euler angles
$\Omega$, $I$, $\Psi$ by:
\begin{equation}
  \begin{aligned}
    q_1 & =\cos\frac{\Omega+\Psi}{2}\cos\frac{I}{2}, & & &
    q_2=\cos\frac{\Omega-\Psi}{2}\sin\frac{I}{2},\\[3pt]
    q_3 & =\sin\frac{\Omega-\Psi}{2}\sin\frac{I}{2}, & & &
    q_4=\sin\frac{\Omega+\Psi}{2}\cos\frac{I}{2}.
  \end{aligned}\label{eq:q1234}
\end{equation}
Note that these parameters satisfy the following relation:
\begin{equation}
  q^2_1+q^2_2+q^2_3+q^2_4=1.
\end{equation}
Taking the time derivatives of Eqs.~(\ref{eq:q1234}), we find
\begin{equation}
  \setlength{\arraycolsep}{5pt}
  \begin{pmatrix}\dot{q}_1\\\dot{q}_2\\\dot{q}_3\\\dot{q}_4\end{pmatrix}=
    \frac{1}{2}\begin{bmatrix}-q_4 & -q_1\tan(I/2) & -q_4\\ -q_3 &
    \hphantom{-}q_2\cot(I/2) & \hphantom{-}q_3\\ \hphantom{-}q_2 &
    \hphantom{-}q_3\cot(I/2) & -q_2\\ \hphantom{-}q_1 &
    -q_4\tan(I/2) &\hphantom{-}q_1\end{bmatrix}
    \begin{pmatrix}\dot{\Omega}\\\dot{I}\\\dot{\Psi}\end{pmatrix}.
\end{equation}
Replacing $\dot{\Omega}$, $\dot{I}$, $\dot{\omega}+\dot{f}$ with the
expressions available in, for example, \citet[][pp.~500,
  501]{Battin_1999}, and using Eq.~(\ref{eq:fict_time}), we arrive at
\begin{equation}
  \begin{pmatrix}q'_1\\q'_2\\q'_3\\q'_4\end{pmatrix}=
  \frac{1}{2}\Big(\frac{h}{r\beta}-\nu'\Big)\begin{pmatrix}-q_4\\\hphantom{-}q_3\\-q_2
  \\\hphantom{-}q_1\end{pmatrix}-\frac{r^2}{2\beta h}F_z
  \begin{pmatrix}\hphantom{-}q_2\cos\nu+q_3\sin\nu\\-q_1\cos\nu+q_4\sin\nu\\-q_4\cos\nu-q_1\sin\nu\\
  \hphantom{-}q_3\cos\nu-q_2\sin\nu\end{pmatrix},
\end{equation}
where $F_z={\bf F}\cdot{\bf e}_z$ and prime denotes
the derivative with respect to $\chi$. By differentiating
Eq.~(\ref{eq:Gauss_eq}) and simplifying the result, we find the
following expression for $\nu'$\footnote{We also used the relation
  \[
  \cos^2\frac{\nu}{2}=\frac{\bigl[r_0U_0\bigl(\frac{1}{2}\chi;\alpha\bigr)
    +\sigma_0U_1\bigl(\frac{1}{2}\chi;\alpha\bigr)\bigr]^2}{rr_0},
  \]
  which can be obtained from Eqs.~(\ref{eq:r2}),~(\ref{eq:p}),
  and~(\ref{eq:Gauss_eq}).}
\begin{multline}
  \beta\nu'=\frac{c}{r}+\frac{r}{c r_0}(rF_r-2\Up)(\alpha
  r_0U_2-\sigma_0U_1)-\frac{\chi c}{\beta r_0}\beta'\\
  -\frac{\mu}{2r_0}\Big[\frac{r}{c}(r_0U_1+\sigma_0U_2)-\frac{c}{\beta^2}U_3\Big]\lambda'.
\end{multline}
The identities involving the universal functions that we used in the
computation of $r'_0$, $\sigma'_0$, $t'_0$, $\nu'$ are reported in
Appendix~\ref{sec:UnivForm}. For negative values of $\TE$, the
presence of secular terms in the derivative of $\nu$ can be avoided
only by setting $\beta=k\sqrt{-2\TE}$, where $k$ is a nonzero
constant. We observe that the four Euler parameters $q_1$, $q_2$,
$q_3$, $q_4$ are constant when the motion is unperturbed. Therefore,
in this case the intermediate frame remains fixed in space.

\subsection{The proposed formulation}
\label{sec:Prop_form}
All that remains to be ready to present our formulation is the
definition of the quantity $\beta$, which was first introduced in
Eq.~(\ref{eq:fict_time}). Several choices are possible in principle;
however it seems natural to set $\beta$ equal to a constant. We take
$\beta=1$ as in the original \citet[][p.~127]{Sundman_1913}
transformation,
\begin{equation}
\frac{\mathrm{d}t}{\mathrm{d}\chi}=r.\label{eq:fict_time2}
\end{equation}
Then, we have $\alpha=\mu\lambda=-2\TE$.

The first four intermediate elements are defined as
\begin{equation}
  \iota_1\coloneqq r_0,\quad\iota_2\coloneqq \sigma_0,\quad\iota_3\coloneqq \alpha,
  \quad\iota_4\coloneqq t_0.
\end{equation}
The remaining four elements are the Euler parameters that represent
the orientation of the intermediate frame:
\begin{equation}
  \iota_5\coloneqq q_1,\quad\iota_6\coloneqq q_2,\quad\iota_7\coloneqq q_3,
  \quad\iota_8\coloneqq q_4.
\end{equation}
The Newtonian equation of motion~(\ref{eq:2bp}) is replaced by the
system of first-order differential equations:
\begin{align}
  \iota'_1 & = -r(rF_r-2\Up)U_1-\frac{\iota'_3}{4}(\iota_1\tilde{U}_2+\iota_2\tilde{U}_3+2\mu U_2^2),
  \label{eq:vop1}\\[3pt]  
  \iota'_2 & = r(rF_r-2\Up)U_0+\frac{\iota'_3}{4}[\iota_1(2\chi+\tilde{U}_1)+\iota_2\tilde{U}_2+
  \mu(\tilde{U}_3-4U_3)],\\[3pt]
  \iota'_3 & = -2\Big(\sigma P_r +hP_{\nu}+r\,\frac{\partial\Up}{\partial t}\Big),\\[3pt]
  \iota'_4 & = r(rF_r-2\Up)U_2-\frac{\iota'_3}{4}[\iota_1(4U_3-\tilde{U}_3)-2\iota_2U_2^2
               -\mu(\tilde{U}_5-8U_5)],\\[3pt]
  \iota'_5 & = -\frac{N}{2}\iota_8-\frac{r^2}{2 h}F_z(\iota_6\cos\nu+\iota_7\sin\nu),\\[3pt]
  \iota'_6 & = \frac{N}{2}\iota_7+\frac{r^2}{2 h}F_z(\iota_5\cos\nu-\iota_8\sin\nu),\\[3pt]
  \iota'_7 & = -\frac{N}{2}\iota_6+\frac{r^2}{2 h}F_z(\iota_8\cos\nu+\iota_5\sin\nu),\\[3pt]
  \iota'_8 & = \frac{N}{2}\iota_5-\frac{r^2}{2 h}F_z(\iota_7\cos\nu-\iota_6\sin\nu),
  \label{eq:vop8}
\end{align}
where
\begin{equation}
  N=\frac{h-c}{r}-\frac{r}{c\iota_1}(rF_r-2\Up)(\iota_1\iota_3U_2-\iota_2U_1)
  +\frac{\iota'_3}{2\iota_1}\Big[\frac{r}{c}(\iota_1U_1+\iota_2U_2)-c\,U_3\Big].
\end{equation}
We want to propagate the position ${\bf r}$ and velocity
$\dot{{\bf r}}$ from some starting epoch $t_*$ to a different
time $t$ by solving the system of
Eqs.~(\ref{eq:vop1})--(\ref{eq:vop8}). The definition of $\iota_i$,
$i=1,...,8$, at $\chi(t_*)=0$ is as follows. For the first four
elements, we have
\begin{equation}
  \iota_{1}=|{\bf r}|,\quad \iota_{2}={\bf r}\cdot\dot{{\bf r}},\quad
  \iota_{3}=\frac{2\mu}{|{\bf r}|}-|\dot{{\bf r}}|^2-2\Up({\bf r},t_*),\quad
  \iota_{4}=t_*.\label{eq:pi14_0}
\end{equation}
Since $\nu=0$ at time $t=t_*$, the intermediate and the LVLH
frames coincide and we can compute
\begin{equation}
  {\bf e}_x=\frac{{\bf r}}{|{\bf r}|},\qquad
  {\bf e}_y={\bf e}_z\times{\bf e}_x,\qquad
  {\bf e}_z=\frac{{\bf r}\times\dot{{\bf r}}}{|{\bf r}\times\dot{{\bf r}}|}.
  \label{eq:exyz}
\end{equation}
The corresponding Euler parameters are obtained by the
formulae~(\ref{eq:q1234}) where $\Psi$ is the argument of latitude,
i.e. $\Psi=\omega+f$. While the sum of $\Omega$ and $\omega+f$ is
defined for any conic, their difference is not. If $I=0$, we can take
$\Omega=0$. In Eqs.~(\ref{eq:pi14_0}), (\ref{eq:exyz}) the quantities
${\bf r}$, $\dot{{\bf r}}$ are referred to the starting epoch.

We solve the initial value problem given by the differential
Eqs.~(\ref{eq:vop1})--(\ref{eq:vop8}) with the initial conditions
computed above by means of a numerical algorithm. At each integration
step, the position ${\bf r}$, velocity $\dot{{\bf r}}$, and
time $t$ can be recovered from the intermediate elements and the
independent variable. We compute
\begin{align}
  r & = \iota_1U_0(\chi;\iota_3)+\iota_2U_1(\chi;\iota_3)+\mu U_2(\chi;\iota_3),\label{eq:erre}\\[3pt]
  \sigma & = \iota_2 U_0(\chi;\iota_3)+(\mu-\iota_1\iota_3)U_1(\chi;\iota_3).\label{eq:sigma}
\end{align}
The components of ${\bf e}_x$, ${\bf e}_y$ in the basis
$\{{\bf e}_1,{\bf e}_2,{\bf e}_3\}$, introduced in
Section~\ref{sec:Mot_ofplane}, are obtained by
\begin{equation}
  {\bf e}_x=\begin{pmatrix} \iota_5^2+\iota_6^2-\iota_7^2-\iota_8^2\\ 2\iota_6\iota_7+2\iota_5\iota_8\\
  2\iota_6\iota_8-2\iota_5\iota_7\end{pmatrix},\qquad
  {\bf e}_y=\begin{pmatrix} 2\iota_6\iota_7-2\iota_5\iota_8\\ \iota_5^2-\iota_6^2+\iota_7^2-\iota_8^2\\
  2\iota_5\iota_6+2\iota_7\iota_8\end{pmatrix}.
\end{equation}
Then, we determine the radial and transverse unit vectors as
\begin{align}
  {\bf e}_r & = \cos\nu\,{\bf e}_x+\sin\nu\,{\bf e}_y,\label{eq:er}\\[3pt]
  {\bf e}_{\nu} & = -\sin\nu\,{\bf e}_x+\cos\nu\,{\bf e}_y,\label{eq:enu}
\end{align}
where
\begin{equation}
  \nu=2\arctan\frac{c\,U_1\left(\frac{1}{2}\chi;\iota_3\right)}{\iota_1
  U_0\left(\frac{1}{2}\chi;\iota_3\right)+\iota_2 U_1\left(\frac{1}{2}\chi;\iota_3\right)}.
  \label{eq:nu}
\end{equation}
The generalised and osculating angular momentum are given from the relations
\begin{equation}
  c^2=\iota_1(2\mu-\iota_1\iota_3)-\iota_2^2,\qquad  h^2=c^2-2r^2\Up({\bf r},t).
  \label{eq:ch}
\end{equation}
Finally, the position and velocity read
\begin{equation}
  {\bf r}=r{\bf e}_r,\qquad \dot{{\bf r}}=\frac{1}{r}(\sigma{\bf e}_r+h{\bf e}_{\nu}),
  \label{eq:rv}
\end{equation}
and the physical time is obtained by
\begin{equation}
  t=\iota_4+\iota_1U_1(\chi;\iota_3)+\iota_2U_2(\chi;\iota_3)+\mu U_3(\chi;\iota_3).
\end{equation}

As expected, our method is affected by the following singularities:
$r=0$, $\iota_1=0$, $h=0$, $c=0$.

\noindent\textit{Remark.} Note that we first compute the potential
$\Up$ and then the osculating angular momentum $h$, which means that
in the proposed method $\Up$ should ideally not depend on the velocity
$\dot{{\bf r}}$. Such limitation can be overcome if $h$ is
regarded as a new state variable. The consequent increase in the
dimension of the system (from 8 to 9) is avoided if one uses the
modification of the Euler parameters suggested by \citet{Lara_2017}.

\section{Orbit determination and uncertainty propagation}
\label{sec:DC}
The topic of orbit determination by means of coordinates different
from the Cartesian ones and an independent variable which is not the
physical time is still quite unexplored. Only very recently,
\citet{RoaPelaez_2017b} have shown in the context of relative motion
that better numerical performance can be achieved with regularised
formulations. Using the results in \citet{Shefer_2007} and
\citet{RoaPelaez_2017a}, we describe in this section how to map the
state-transition matrix (STM) of the intermediate elements at some
fictitious time $\chi$ to the classic STM in Cartesian coordinates at
the corresponding time $t$.

Let us denote by ${\bm\iota}\in\mathbb{R}^8$ the column vector of the
intermediate elements, i.e. ${\bm\iota}=(\iota_1,\ldots,\iota_8)^T$,
so that we can write Eqs.~(\ref{eq:vop1})--(\ref{eq:vop8}) in the
compact form
\begin{equation}
  \frac{\mathrm{d}{\bm\iota}}{\mathrm{d}\chi}={\bf f}(\chi,{\bm\iota}).\label{eq:dy_dchi}
\end{equation}
For the solution ${\bm\iota}(\chi,{\bm\iota}_0)$ of~(\ref{eq:dy_dchi})
with the initial condition ${\bm\iota}_0={\bm\iota}(0)$ we introduce
the state-transition matrix
\begin{equation}
  A(\chi,{\bm\iota}_0)=\frac{\partial{{\bm\iota}}}{\partial{{\bm\iota}_0}}(\chi,{\bm\iota}_0).
\end{equation}
The matrix $A$ satisfies the Cauchy problem
\begin{equation}
  \frac{\partial{A}}{\partial\chi}=\frac{\partial{\bf f}}{\partial{\bm\iota}}
  ({\bm\iota}(\chi,{\bm\iota}_0))A,\quad
  A(0,{\bm\iota}_0)=I_d,\label{eq:ve}
\end{equation}
where $I_d$ is the $9\times 9$ identity matrix. The
solution of the differential equation in~(\ref{eq:ve}), which is
called variational equation, is computed numerically together with the
solution of Eq.~(\ref{eq:dy_dchi}).

Let ${\bf x}\in\mathbb{R}^6$ be the (column) vector with
components given by the coordinates of the position ${\bf r}$ and
velocity $\dot{{\bf r}}$ of a space object in a suitable
reference frame. Suppose we apply the method known as
\textit{differential corrections} \citep[for more details,
  see][chap.~5]{MilaniGronchi_2010} to determine
${\bf x}_0={\bf x}(t_*)$, from a set of observations
collected at times $t_1<t_2<\ldots<t_m$.\footnote{The time $t_*$ is
  usually chosen as the average of the observation times.} In each
iteration of this method, we need the state-transition matrix
\begin{equation}
  S(t,{\bf x}_0)=\frac{\partial{{\bf x}}}{\partial{{\bf x}_0}}(t,{\bf x}_0)
\end{equation}
at $t_i$, $i=1,\ldots,m$. We can calculate
$S(t,{\bf x}_0)$ from
$A(\chi,{\bm\iota}_0)$ through the formula (we set $\chi=0$
at $t=t_*$)
\begin{equation}
  S=J\,\tilde{A}\,J_0,\label{eq:STMfromA}
\end{equation}
where
\begin{equation}
  J=\frac{\partial{\bf x}}{\partial{\bm\iota}},\qquad
  J_0=\frac{\partial{\bm\iota}_0}{\partial{\bf x}_0},\qquad
  \tilde{A}=A-\frac{1}{r}{\bf f}\frac{\partial{t}}{\partial{\bm\iota}_0}.
\end{equation}

Starting from a first guess of ${\bf x}_0$, the iterative method
converges to a nominal solution. Moreover, the associated covariance
matrix ${\Gamma}_{{\bf x}_0}$ is known. One may want to
propagate ${\Gamma}_{{\bf x}_0}$ to
${\Gamma}_{{\bf x}}$ at time $t\neq t_*$. Linear propagation
is in many cases acceptable, and it can be made more efficient if we
use orbital elements \citep{Junkins_etal_1996}. First, the matrix
${\Gamma}_{{\bf x}_0}$ is transformed to
${\Gamma}_{{\bm\iota}_0}=J_0\,{\Gamma}_{{\bf x}_0}\,J_0^T$. Then,
the covariance matrix at time $t$ is obtained by
\begin{equation}
  {\Gamma}_{{\bm\iota}}=\tilde{A}\,{\Gamma}_{{\bm\iota}_0}\,\tilde{A}^T.
\end{equation}
Finally, we apply the conversion ${\Gamma}_{\bf
  x}=J\,{\Gamma}_{\bm\iota}\,J^T$. The explicit expressions of the
Jacobian matrices $J$, $J_0$ are given in the following sections, and
of the matrix $\partial{{\bf f}}/\partial{\bm\iota}$ in
Appendix~\ref{sec:dfdi}.

\subsection{Partial derivatives of position and velocity with respect to intermediate elements}
\label{sec:drv_di}
Let us first give the expression of the Jacobian matrix
$\partial{{\bf r}}/\partial{\bm\iota}$. From the first relation
in~(\ref{eq:rv}) and using Eqs.~(\ref{eq:er}),~(\ref{eq:enu}), we have
\begin{equation}
  \frac{\partial{{\bf r}}}{\partial{\bm\iota}}={\bf e}_r\frac{\partial r}{\partial{\bm\iota}}+
  r\frac{\partial{{\bf e}}_r}{\partial{\bm\iota}},
\end{equation}
where
\begin{equation}
  \frac{\partial{{\bf e}}_r}{\partial{\bm\iota}}={\bf e}_{\nu}\frac{\partial\nu}{\partial{\bm\iota}}+
  \frac{\partial{{\bf e}}_x}{\partial{\bm\iota}}\cos\nu+\frac{\partial{{\bf e}}_y}{\partial{\bm\iota}}
  \sin\nu.
\end{equation}
Differentiation of the expressions for $r$, $\nu$, $c$
in~(\ref{eq:erre}),~(\ref{eq:nu}),~(\ref{eq:ch}) yields
\begin{align*}
  \frac{\partial r}{\partial{\bm\iota}} & = \Bigl(U_0,\,U_1,\,\frac{1}{2}[\iota_2U_3+2\mu U_4
  -\chi(t-\iota_4)],\,{\bf 0}_5\Bigr),\\[3pt]
  r\iota_1\frac{\partial\nu}{\partial{\bm\iota}} & = (\iota_1U_1+\iota_2U_2)\frac{\partial c}
  {\partial{\bm\iota}}-c\Bigl(U_1,\,U_2,\,-\frac{\iota_1U_3}{2},\,{\bf 0}_5\Bigr),\\[3pt]
  \frac{\partial c}{\partial{\bm\iota}} & = \frac{1}{c}\Bigl(\mu-\iota_1\iota_3,\,-\iota_2,\,
  -\frac{\iota_1^2}{2},\,{\bf 0}_5\Bigr),
\end{align*}
where ${\bf 0}_5\in\mathbb{R}^5$ is a row vector having null
entries. The unit vectors ${\bf e}_x$, ${\bf e}_y$ are
functions only of the four Euler parameters, and
\begin{align*}
  \frac{\partial{{\bf e}}_x}{\partial(\iota_5,\iota_6,\iota_7,\iota_8)} & = 2
  \arraycolsep=4.4pt
  \left(\hspace{-0.1cm}\begin{array}{rrrr}
     \hphantom{-}\iota_5 & \iota_6 &            -\iota_7 &            -\iota_8\\
     \hphantom{-}\iota_8 & \iota_7 & \hphantom{-}\iota_6 & \hphantom{-}\iota_5\\
                -\iota_7 & \iota_8 &            -\iota_5 & \hphantom{-}\iota_6
  \end{array}\hspace{-0.1cm}\right),\\[3pt]
  \frac{\partial{{\bf e}}_y}{\partial(\iota_5,\iota_6,\iota_7,\iota_8)} & = 2
  \arraycolsep=4.4pt
  \left(\hspace{-0.1cm}\begin{array}{rrrr}
                -\iota_8 & \hphantom{-}\iota_7 & \iota_6 &            -\iota_5\\
     \hphantom{-}\iota_5 &            -\iota_6 & \iota_7 &            -\iota_8\\
     \hphantom{-}\iota_6 & \hphantom{-}\iota_5 & \iota_8 & \hphantom{-}\iota_7
  \end{array}\hspace{-0.1cm}\right).
\end{align*}
Then, we deal with the Jacobian matrix
$\partial{\dot{{\bf r}}}/\partial{\bm\iota}$. From the second
relation in~(\ref{eq:rv}) and Eq.~(\ref{eq:enu}), we get
\begin{equation}
  \frac{\partial{\dot{{\bf r}}}}{\partial{\bm\iota}}=\frac{1}{r}\Bigl({\bf e}_r
  \frac{\partial\sigma}{\partial{\bm\iota}}+{\bf e}_{\nu}\frac{\partial h}{\partial{\bm\iota}}
  +\sigma\frac{\partial{{\bf e}}_r}{\partial{\bm\iota}}+h\frac{\partial{{\bf e}}_{\nu}}
  {\partial{\bm\iota}}-{\dot{{\bf r}}}\frac{\partial r}{\partial{\bm\iota}}\Bigr),
\end{equation}
where
\begin{equation}
  \frac{\partial{{\bf e}}_{\nu}}{\partial{\bm\iota}}=-{\bf e}_r\frac{\partial\nu}{\partial{\bm\iota}}-
  \frac{\partial{{\bf e}}_x}{\partial{\bm\iota}}\sin\nu+\frac{\partial{{\bf e}}_y}{\partial{\bm\iota}}
  \cos\nu.
\end{equation}
Using the expressions of $\sigma$, $h$
in~(\ref{eq:sigma}),~(\ref{eq:ch}), we can compute
\begin{align*}
  \frac{\partial\sigma}{\partial{\bm\iota}} & = \Bigl(-\iota_3U_1,\,U_0,\,
  \frac{1}{2}(\mu U_3-\iota_1U_1-\chi r),\,{\bf 0}_5\Bigr),\\[3pt]
  \frac{\partial h}{\partial{\bm\iota}} & = \frac{1}{h}\Bigl(\mu-\iota_1\iota_3-2rU_0\Up,
  \,-\iota_2-2rU_1\Up,-\frac{\iota^2_1}{2}-2r\Up\frac{\partial r}{\partial \iota_3},\,{\bf 0}_5\Bigr)
  -\frac{r^2}{h}\frac{\partial\Up}{\partial{\bm\iota}}.
\end{align*}
The vector $\partial\Up/\partial{\bm\iota}$ is obtained as shown in
Eq.~(\ref{eq:dU_di}).

\noindent\textit{Remark.} We observe that when $\TE<0$, secular terms
appear only in 6 out of the 48 components of the matrix of the partial
derivatives of position and velocity with respect to intermediate
elements, denoted by $J$. As observed, for example, by
\citet{BrouckeCefola_1972}, this is a remarkable advantage over the
universal variables, in which the fundamental matrix contains secular
terms in all 36 elements.

\subsection{Partial derivatives of intermediate elements with respect to position
  and velocity at the initial time}
\label{sec:di_drv}
All the quantities of this section are referred to the initial time
$t_*$ of propagation. For the first four intermediate elements, a
straightforward computation from relations~(\ref{eq:pi14_0}) yields
\begin{equation}
\begin{aligned}
  \frac{\partial\iota_{1}}{\partial{{\bf r}}} & = {\bf e}_x^T, & & & \frac{\partial\iota_{1}}
  {\partial{\dot{{\bf r}}}} & = {\bf 0}_3,\\[3pt]
  \frac{\partial\iota_{2}}{\partial{{\bf r}}} & = {\dot{{\bf r}}}^T, & & & \frac{\partial\iota_{2}}
  {\partial{\dot{{\bf r}}}} & = {{\bf r}}^T,\\[3pt]
  \frac{\partial\iota_{3}}{\partial{{\bf r}}} & = -2\Bigl(\frac{\mu}{r^2}{{\bf e}}_x^T+\frac{\partial\Up}
  {\partial{{\bf r}}}\Bigr), & & & \frac{\partial\iota_{3}}{\partial{\dot{{\bf r}}}} & =
  -2\dot{{\bf r}}^T,\\[3pt]
  \frac{\partial\iota_{4}}{\partial{{\bf r}}} & = {\bf 0}_3, & & & \frac{\partial\iota_{4}}
  {\partial{\dot{{\bf r}}}} & = {\bf 0}_3,
\end{aligned}
\end{equation}
where ${\bf 0}_3=(0,\,0,\,0)$.

At the initial time the intermediate and the LVLH frames coincide (see
Eqs.~\ref{eq:exyz}). The partial derivatives of the Euler parameters
take the following simple expressions which are obtained as shown in
Appendix~\ref{sec:dEP_drv}:
\begin{equation}
\begin{split}
  \frac{\partial \iota_5}{\partial{{\bf r}}} & = \frac{1}{2 r}[(\iota_7+
  \iota_6\upsilon){{\bf e}}_z^T-\iota_8\,{{\bf e}}_y^T],\\[3pt]
  \frac{\partial \iota_6}{\partial{{\bf r}}} & = \frac{1}{2 r}[(\iota_8-
  \iota_5\upsilon){{\bf e}}_z^T+\iota_7\,{{\bf e}}_y^T],\\[3pt]
  \frac{\partial \iota_7}{\partial{{\bf r}}} & = -\frac{1}{2 r}[(\iota_8\upsilon+
  \iota_5){{\bf e}}_z^T+\iota_6\,{{\bf e}}_y^T],\\[3pt]
  \frac{\partial \iota_8}{\partial{{\bf r}}} & = \frac{1}{2 r}[(\iota_7\upsilon-
  \iota_6){{\bf e}}_z^T+\iota_5\,{{\bf e}}_y^T],
\end{split}\label{eq:dq58_dr}
\end{equation}
where $\upsilon=\sigma/h$, and 
\begin{equation}
\begin{aligned}
  \frac{\partial \iota_5}{\partial{\dot{{\bf r}}}} & = -\frac{r}{2 h}\iota_6\,{{\bf e}}_z^T, & & &
  \frac{\partial \iota_6}{\partial{\dot{{\bf r}}}} & = \frac{r}{2 h}\iota_5\,{{\bf e}}_z^T,\\[3pt]
  \frac{\partial \iota_7}{\partial{\dot{{\bf r}}}} & = \frac{r}{2 h}\iota_8\,{{\bf e}}_z^T, & & &
  \frac{\partial \iota_8}{\partial{\dot{{\bf r}}}} & = -\frac{r}{2 h}\iota_7\,{{\bf e}}_z^T.
\end{aligned}\label{eq:dq58_dv}
\end{equation}

\section{Numerical tests}
\label{sec:num_res}
We present two numerical tests to have a taste of the performance of
the intermediate elements compared to other methods existing in the
literature. In particular, we choose Cowell's method
\citep[][p.~447]{Battin_1999}, the modified equinoctial elements
\citep{Walker_etal_1985}, Dromo \citep{Pelaez_etal_2007}, the natural
elements derived by \citet{Burdet_1968}, and the regular KS-based
elements published by \citet{Bond_1974}. Hereafter, we will refer to
them as Cowell, ModEq, Dromo, Nat--Burdet, and KS--Bond,
respectively. Some relevant features of these formulations and the
intermediate elements are given in Table~\ref{Tab:Formuls}. We use the
label ``New'' to refer to the formulation described in
Section~\ref{sec:Prop_form}.


Performance is assessed by analysing accuracy and speed of each
special perturbation method. The accuracy is measured by propagating
the orbit forward until the final time, reversing the integration back
to the initial time, and computing the error as the difference between
the final state (position and velocity) and the initial one. The speed
of a particular integration is determined by the number of times the
integrator calls the force function. This metric is preferred over the
actual runtime because it is machine and implementation
independent. In real scenarios, evaluating the perturbation forces is
computationally more expensive than the rest of operations required to
calculate the right-hand side of the differential equations. Thus, the
additional cost per function call that one has to pay for using a
formulation that is more sophisticated than Cowell is usually
negligible. The selected integrator is the standard variable step
Runge--Kutta 4(5) implemented in Matlab's \texttt{ode45}
function. Performance curves are generated by changing the relative
tolerance from $10^{-6}$ to $10^{-13}$. We note that Nat--Burdet,
KS--Bond, and the new elements make use of Stumpff's functions. The
series are evaluated via recursive formulae implementing an
argument-reduction technique to ensure convergence, as indicated by
\citet[][sect.~6.9]{Danby_Book} and \citet[Appendix
  1]{RoaPelaez_2017b}.

\begin{table}
  \centering
  \caption{Formulations compared in the numerical tests (labels are
    explained in the text above). The quantities $t$, $r$, $h$ are time, the
    orbital radius, and the osculating angular momentum. Prime denotes
    the derivative with respect to the independent variable. For each
    formulation, we specify the adopted independent variable, the
    dimension of the state vector, the number of orbital elements
    (i.e. constants of the unperturbed motion) among the state
    variables, and if a time element is included.}
  \label{Tab:Formuls}
  \begin{tabular}{lccccc}
    \toprule
    Formulations & Indep. variable & Dim. & Elements & Time el.\tabularnewline
    \midrule
    Cowell & $t$ & 6 & 0 & no\tabularnewline
    ModEq & $t$ & 6 & 5 & no\tabularnewline
    Dromo & $t'=r^2/h$ & 8 & 7 & no\tabularnewline
    Nat--Burdet & $t'=r$ & 11 & 11 & yes\tabularnewline
    KS--Bond & $t'=r$ & 10 & 10 & yes\tabularnewline
    New & $t'=r$ & 8 & 8 & yes\tabularnewline
    \bottomrule
  \end{tabular}
\end{table}

\subsection{The hyperbolic comet C/2003~T4 (LINEAR)}
\label{Sec:2003t4}
We consider the orbit of the hyperbolic comet C/2003~T4 (LINEAR),
defined by its osculating elements in
Table~\ref{Tab:orbit_comet_C2003T4}. Non-gravitational forces are
modelled following \citet{Marsden_etal_1973}: we use the coefficients
$A_1=1.0592\times10^{-7}$~au\,d$^{-2}$,
$A_2=8.1043\times10^{-10}$~au\,d$^{-2}$, and
$A_3=3.2073\times10^{-9}$~au\,d$^{-2}$ for the radial, transverse, and
normal components, respectively (d stands for day). Although the
resulting acceleration is small, obviating this effect results in a
non-negligible separation of approximately 0.05~au at the final
epoch. Gravitational perturbations are given by the attraction of the
outer planets (Jupiter through Neptune). Their positions are retrieved
from the DE431 ephemeris.

\begin{table*}
  \centering
  \caption{Osculating elements of C/2003~T4 (LINEAR) at epoch
    JD~2453296.5 (2004 October~18) TDB (Barycentric Dynamical Time). Orbit
    solution JPL~132. They are the eccentricity ($e$), perihelion
    distance ($q$, in astronomical units), time of perihelion passage
    ($t_p$), inclination ($I$), longitude of the node ($\Omega$), and
    argument of perihelion ($\omega$). Angles are in degrees.}
    \label{Tab:orbit_comet_C2003T4}
  \begin{tabular}{cccccc}
    \toprule
    $e$ & $q$ (au) & $t_p$ (TDB) & $I$ & $\Omega$ & $\omega$\\
    \midrule
    1.0005 & 0.8498 & 2453464.16 & 86.7612 & 93.9029 & 181.6795\\
    \bottomrule
  \end{tabular}
\end{table*}

\begin{figure}[h]
  \centering
  \includegraphics[width=0.97\linewidth]{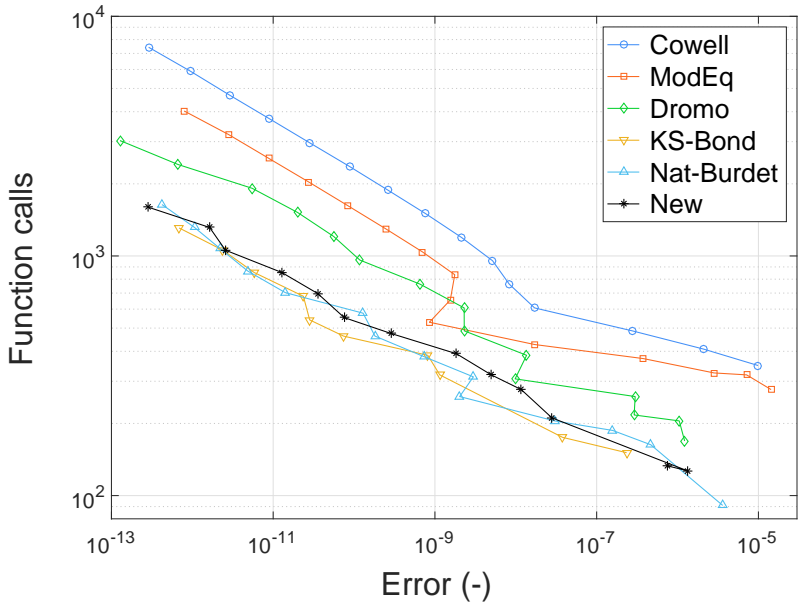}
  \caption{Performance of different propagation methods when
    integrating the orbit of comet C/2003~T4. The state error is
    normalised using the heliocentric distance of the comet at the
    initial time as the unit of length and the unit of time is chosen 
    so that the gravitational parameter is normalised to unity.}
  \label{Fig:err_comet_C2003T4}
\end{figure}

The orbit reported in Table~\ref{Tab:orbit_comet_C2003T4} is
propagated for 10~years starting on 2000 April~3, that is 5~years
before perihelion passage. Figure~\ref{Fig:err_comet_C2003T4} displays
the performance curves for each of the selected formulations. Since
external perturbations are weak in this example, using the modified
equinoctial elements instead of Cartesian coordinates reduces the
number of function calls by approximately a factor of two for the same
accuracy. Time is the independent variable also for ModEq, which means
that the performance gains with respect to Cowell are only due to the
use of slowly varying variables. In Dromo, the fictitious time behaves
like the true anomaly when the motion is Keplerian, and it results in
this method being four and two times faster than Cowell and ModEq,
respectively. The improvement comes from a more efficient step size
control, specifically during pericentre passage. The methods
Nat--Burdet, KS--Bond, and the intermediate elements presented in this
paper all rely on a (first-order) Sundman time transformation. Thus,
the independent variable evolves like the hyperbolic anomaly, which
naturally optimises the discretisation of hyperbolic orbits. These
three formulations exhibit the best performance: they are almost one
order of magnitude faster than the integration in Cartesian
coordinates in this particular example. Only when the comet is close
to perihelion, the step size is slightly reduced, although the
reduction is only by a factor of two compared to the two order of
magnitude reduction observed when time is the independent
variable. The more efficient discretisation of the orbit around
perihelion produces the aforementioned improvements in performance.

\begin{figure}[t]
  \centering
  \includegraphics[width=0.97\linewidth]{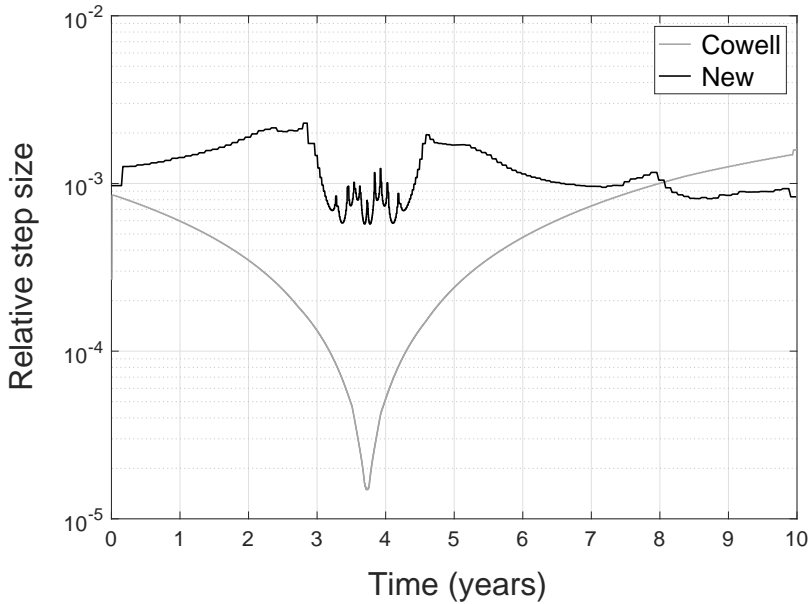}
  \caption{Evolution of the integration step size during the
    propagation of the orbit of comet C/2003~T4. The close encounter
    with the Sun causes a strong reduction of the step for Cowell,
    while it  has almost no effect on the new method.
    \label{Fig:stepsize_comet}}
\end{figure}

The positive effect of the analytic step size adaption, observed
thanks to introducing a modified time variable, becomes apparent in
Fig.~\ref{Fig:stepsize_comet}. This figure compares the evolution of
the time step when Cowell's method and the new elements are used to
propagate the orbit. Cowell sequentially reduces the step size as the
comet approaches the perihelion. This is required to meet the
integration tolerance as the velocity increases and the problem
becomes more sensitive to small deviations. Conversely, the length of
the integration step shows a small variation when the orbit is
propagated by the intermediate elements. After the comet passes the
closest approach with the Sun and moves away from it along the
outgoing asymptote, the step size increases again for Cowell, while the
velocity decreases. Finally, we observe that far enough from the
perihelion, it becomes comparable to that of the new formulation.

\subsection{The comet C/1985~K1 (Machholz)}
\begin{figure}
  \centering
  \includegraphics[width=0.97\linewidth]{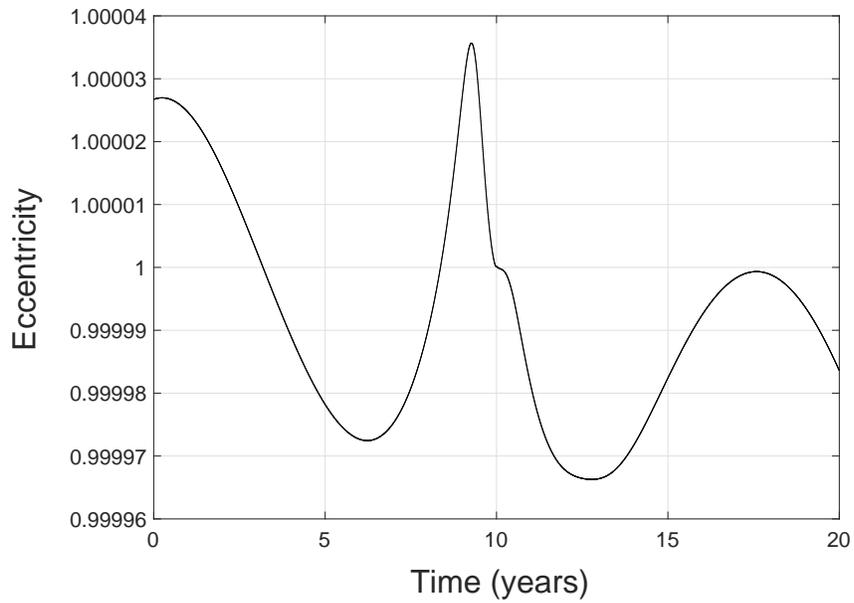}
  \caption{Evolution of the eccentricity of comet C/1985~K1.
    \label{Fig:eccev_1985k1}}
\end{figure}

Gravitational perturbations from the outer planets cause the orbit of
comet C/1985~K1 to transition between hyperbolic and elliptic
repeatedly, as shown in Fig.~\ref{Fig:eccev_1985k1}. This behaviour is
ideal for testing how a uniform formulation handles different orbital
regimes. The orbit is initially hyperbolic (see the osculating orbital
elements listed in Table~\ref{Tab:orbit_comet_1985k1}), the
eccentricity decreases as the comet approaches perihelion temporarily
becoming less than unity, then increasing to produce a hyperbolic
orbit at perihelion. The orbit is propagated for 20~years, starting
10~years before perihelion.

\begin{figure}
  \centering
  \includegraphics[width=0.97\linewidth]{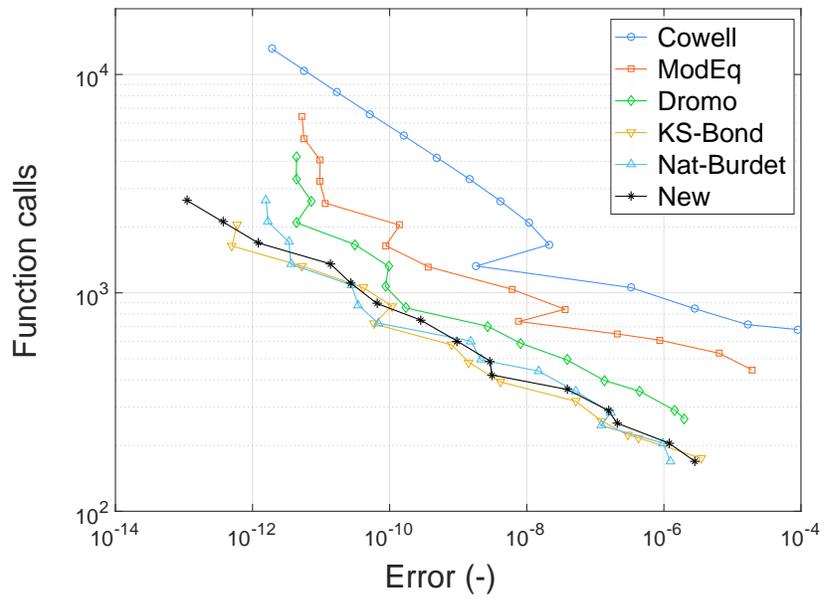}
  \caption{Performance of different propagation methods when
    integrating the orbit of comet C/1985~K1.}
  \label{Fig:err_comet_1985k1}
\end{figure}

To propagate the orbit of comet C/1985~K1, we resort to the numerical
setup described in Section~\ref{Sec:2003t4} except that
non-gravitational forces are not included.
Figure~\ref{Fig:err_comet_1985k1} compares the performance of the
selected formulations. Although Cowell's method does not depend
explicitly on the type of orbit, its overall performance is affected
by the fact that using the physical time as independent variable
results in an inefficient discretisation of the orbit. The orbital
elements evolve slowly over time, and consequently, employing the
modified equinoctial elements instead of Cartesian coordinates
produces a substantial improvement in performance. Next, Dromo
replaces the physical time with the true anomaly and the performance
gain observed in Fig.~\ref{Fig:err_comet_1985k1} relative to ModEq is
due to the analytic step size adaption implicit in the change of the
independent variable. The best behaviour is shown by the formulations
relying on the Sundman transformation~(\ref{eq:fict_time2}): KS--Bond,
Nat--Burdet, and the new elements. The performance of these three
formulations is similar, with the intermediate elements being slightly
more accurate for small integration tolerances. The new formulation is
capable of transitioning between orbital regimes without singularities
or accuracy losses.

\begin{table*}
  \centering
  \caption{Osculating elements of C/1985~K1 (Machholz) at epoch
    JD~2442592.7 (1975 June~29) TDB (Barycentric Dynamical Time). Orbit
    solution from 2008~SAO Comet Catalog. Angles are in degrees.}
    \label{Tab:orbit_comet_1985k1}
  \begin{tabular}{cccccc}
    \toprule
    $e$ & $q$ (au) & $t_p$ (TDB) & $I$ & $\Omega$ & $\omega$\\
    \midrule
    1.000026 & 0.1085 & 2446245.24 & 16.0812 & 198.2520	& 271.7063\\
    \bottomrule
  \end{tabular}
\end{table*}

\section{Conclusions}
\label{sec:thend}
Uniform and regular orbital elements for propagating the motion of a
celestial body in the perturbed two-body problem have been developed
so far only from the KS and Sperling's regularisations. In both
methods, the variables from which the elements are generated contain
information on the radial distance and the orientation of the radial
unit vector. By contrast, if the separate evolution of these two
quantities is considered, as in the Burdet--Ferr\'andiz (BF)
regularisation, an intrinsic singularity arises for $r=0$.

In this work, a formulation that consists of eight orbital elements is
presented. We derive them following the spirit of BF decomposition,
but we introduce a new time variable $\chi$ through a transformation
of the Sundman type, instead of using Eq.~(\ref{eq:phi}) as in the BF
regularisation. First, for the radial displacement $r$, we find a
second-order linear differential equation with constant coefficients
(as expected). The two intermediate elements ($r_0$, $\sigma_0$) that
stem from its solution correspond to the values of $r$ and its
derivative with respect to $\chi$ at the epoch $t_0$. We note that
secular terms affect their derivatives for negative values of the
total energy, and we identified the reason. This drawback is common to
all the universal formulations based on orbital elements (see
Introduction) and can be avoided only at the price of losing
universality. Then, following a more geometric approach, we define an
intermediate reference frame whose evolution keeps track of the
orientation of the orbital plane and of a reference direction on
it. From such direction, the position of the particle is obtained by a
counterclockwise rotation of $\nu$ (Eq.~\ref{eq:nu}) about the angular
momentum vector. The angle $\nu$ is determined by $\chi$, the total
energy, $r_0$, $\sigma_0$, and when $\Up=0$, it corresponds to the
difference between the true anomalies at times $t$ and $t_0$ (measured
from the pericentre of the osculating conic at time $t$). The total
energy (multiplied by $-2$), the time element $t_0$, and four Euler
parameters associated with the intermediate frame are the remaining
six intermediate elements. The resulting set is uniform, but not
universal because it does not work when the angular momentum is
zero. An alternative formulation which is completely free of secular
terms is also presented.

In addition to a pure theoretical interest in alternative variables
for new special perturbation methods, we are concerned with their
practical utility. Numerical tests performed by the authors and others
corroborate our belief that they can be much more accurate and faster
than the classic computation of orbits with Cartesian coordinates. We
are aware that the implementation becomes more difficult, because the
independent variable is not the physical time and the conversion
between position and velocity and the new quantities involves
complicated expressions. Therefore, in order to encourage the reader
to code the proposed method, we have reported all the necessary
formulae for propagating initial conditions and computing orbits from
observations by means of the differential corrections method.

Finally, we tested the performance of the intermediate elements by
evaluating their accuracy and computational speed with respect to
several other methods. For this purpose, we propagated the orbits of
the hyperbolic comet (with eccentricity almost equal to 1) C/2003~T4
(LINEAR) and of the comet C/1985~K1 (Machholz) whose eccentricity
fluctuates around 1. The new elements and other two universal
formulations, which rely on Sundman's time transformation and include
a time element, substantially outperform Cowell's method. In the case
of C/1985~K1, the intermediate elements reach the highest accuracy with
a relatively small computational cost. Finally, we also checked that
secular terms do not affect the performance shown by the proposed
formulation for propagation times in the order of centuries/thousands
of years.

\section{Acknowledgements}
The author G. Ba\`u acknowledges the project MIUR-PRIN 20178CJA2B
titled ``New frontiers of Celestial Mechanics: theory and
applications''. Part of this work was carried out at the Jet
Propulsion Laboratory, California Institute of Technology, under a
contract with the National Aeronautics and Space Administration.  We
also thank Z. Kne\v{z}evi{\'c} for providing us with the correct
reference to the paper of M. Milankovi{\'c}, and the reviewers for
their useful comments.



\appendix
\section{Avoiding secular terms, alternative formulation}
\label{sec:nosec}
When the total energy is negative ($\TE<0$), secular terms that appear
in the derivatives of the intermediate elements
(Eqs.~\ref{eq:vop1}--\ref{eq:vop8}) may deteriorate the accuracy of
the predicted state in long-term propagations. The proposed
formulation can be modified in order to overcome this drawback. The
idea is inspired by the regularised methods presented in
\citet[][chap.~1]{Stiefel_etal_1967}.

By setting $\beta=1$, we can
rewrite Eq.~(\ref{eq:d2r_dchi2}) as
\begin{equation*}
  r''=2\bar{\TE} r+\mu+r(rF_{r}+2\TE_K-2\bar{\TE}),
\end{equation*}
where $\TE_K=\TE-\Up$ is the Keplerian energy and $\bar{\TE}$ is the
value taken by $\TE$ at the initial time of the propagation. The
quantity $\bar{\TE}$ is a constant which is fixed by the initial
position and velocity of the particle. Let us introduce
  \begin{equation*}
    \bar{\alpha}=-2\bar{\TE},\qquad\alpha=-2\TE,\qquad\delta\alpha=\bar{\alpha}-\alpha.
  \end{equation*}
The solution of $r''=2\bar{\TE}
r+\mu$ is given by
\begin{equation}
  r=r_0u_0(\chi;\bar{\alpha})+\sigma_0u_1(\chi;\bar{\alpha})+\mu u_2(\chi;\bar{\alpha}),
  \label{eq:r_new}
\end{equation}
where we have introduced the universal functions (see
Eq.~\ref{eq:Un}):
\begin{equation*}
  u_{n}(\chi;\bar{\alpha})=\chi^n\sum_{k=0}^{\infty}(-1)^{k}
  \frac{(\bar{\alpha}\chi^2)^k}{(n+2k)!},\quad n\in\mathbb{N}.
  \label{eq:un}
\end{equation*}
The derivatives of $u_n$ with respect to $\chi$ do not contain secular
terms. Following the same steps as in Sections~\ref{sec:VOP_rsl0}
and~\ref{sec:Kep_eq}, we obtain
\begin{align*}
  r_0' & = -ru_1(rF_r-2\Up+\delta\alpha),\\[2pt]
  \sigma_0' & = ru_0(rF_r-2\Up+\delta\alpha),\\[2pt]
  t_0' & = ru_2(rF_r-2\Up+\delta\alpha).
\end{align*}
Then, by substituting in Eq.~(\ref{eq:nu_int}) the expression of $r$
in~(\ref{eq:r_new}) we arrive at the Gaussian equation
\begin{equation*}
  \tan\frac{\nu}{2}=\frac{c\,u_1\left(\frac{1}{2}\chi;\bar{\alpha}\right)}{r_0
    u_0\left(\frac{1}{2}\chi;\bar{\alpha}\right)+\sigma_0
    u_1\left(\frac{1}{2}\chi; \bar{\alpha}\right)}.
\end{equation*}
Differentiation with respect to $\chi$ yields
\begin{equation*}
  \nu'=\frac{2}{d}\Bigl[\frac{c}{r}(d-r_0)+\frac{r}{c}(rF_r-2\Up)
    (\bar{\alpha}r_0u_2-\sigma_0u_1)-\frac{\alpha'r}{2c}(r_0u_1+\sigma_0u_2)\Bigr],
\end{equation*}
where
\begin{equation*}
  d=2r_0+r\delta\alpha\,u_2.
\end{equation*}

The expressions of $r_0'$, $\sigma_0'$, $t_0'$, $\nu'$ reported above
do not contain secular terms for $\bar{\TE}<0$. Thus, we can select
$r_0$, $\sigma_0$, $\alpha$, $t_0$ and four Euler parameters exactly
as we did in Section~\ref{sec:Prop_form}, for the elements of a
formulation of the perturbed two-body problem, which will be free of
secular terms. The case $d=0$ does not introduce additional
singularities to those affecting the intermediate elements (see
Section~\ref{sec:Prop_form}).\footnote{In fact, it can be shown that
  $rd=2c^2u_1^2\bigl(\frac{1}{2}\chi;\bar{\alpha}\bigr)/\sin^2\frac{\nu}{2}$.}
Finally, we note that the same relation as in~(\ref{eq:ch}) for $c^2$
does not hold anymore, and we have to use instead
\begin{equation*}
  c^2=r_0(2\mu-r_0\bar{\alpha})-\sigma_0^2+\delta\alpha\,r^2.
\end{equation*}

\section{Partial derivatives of $\iota_5$, $\iota_6$, $\iota_7$, $\iota_8$ with respect to
  position and velocity at the initial time}
\label{sec:dEP_drv}
We show a possible way of deriving formulae~(\ref{eq:dq58_dr})
and~(\ref{eq:dq58_dv}). We recall that ${\bf e}_r$, ${\bf e}_{\nu}$,
${\bf e}_z$ are the unit vectors of the LVLH reference frame (see
Eqs.~{\ref{eq:ernuz}}). From the following relation for the angular
momentum vector
\begin{equation*}
  {\bf r}\times{\dot{{\bf r}}}=h{\bf e}_z,
\end{equation*}
we obtain
\begin{equation}
  h\frac{\partial{{\bf e}}_z}{\partial{{\bf r}}}=
  V-{\bf e}_z\frac{\partial h}{\partial{{\bf r}}},\qquad
  h\frac{\partial{{\bf e}}_z}{\partial{\dot{{\bf r}}}}=
  R-{\bf e}_z\frac{\partial h}{\partial{\dot{{\bf r}}}},
  \label{eq:dez_drdv}
\end{equation}
where $V$, $R$ are the skew-symmetric matrices
defined by
\begin{align*}
  V(1,2) & =v_3, & V(1,3) & =-v_2, & V(2,3) & =v_1,\\[2pt]
  R(1,2) & =-r_3, & R(1,3) & =r_2, & R(2,3) & =-r_1,
\end{align*}
with $r_i={\bf r}\cdot{\bf e}_i$,
$v_i={\dot{{\bf r}}}\cdot{\bf e}_i$, $i=1,2,3$, and
\begin{equation*}
  \frac{\partial h}{\partial{{\bf r}}}={{\bf e}}_z^TV,\qquad
  \frac{\partial h}{\partial{\dot{{\bf r}}}}={{\bf e}}_z^TR.
\end{equation*}
By inserting in Eqs.~(\ref{eq:dez_drdv}) the expression of
${\bf e}_z$ as a function of $I$, $\Omega$, that is
\begin{equation*}
  {\bf e}_z={\bf e}_1\sin\Omega\sin I-{\bf e}_2\cos\Omega\sin I+{\bf e}_3\cos I,
\end{equation*}
we find
\begin{align}
  \frac{\partial\Omega}{\partial{{\bf r}}} & = -\frac{1}{p\sin I}(\cos L+e\cos\omega){{\bf e}}_z^T, &
  \frac{\partial\Omega}{\partial{\dot{{\bf r}}}} & = \frac{r\sin L}{h\sin I}\,{{\bf e}}_z^T,
  \label{eq:dOM_drv}\\[2pt]
  \frac{\partial I}{\partial{{\bf r}}} & = \frac{1}{p}(\sin L+e\sin\omega){{\bf e}}_z^T, &
  \frac{\partial I}{\partial{\dot{{\bf r}}}} & = \frac{r}{h}\cos L\,{{\bf e}}_z^T,
  \label{eq:dI_drv}
\end{align}
where $L=\omega+f$ is the argument of latitude.

Then, from the relation
\begin{equation*}
  \cos L=({\bf e}_r\cdot{\bf e}_1)\cos\Omega+({\bf e}_r\cdot{\bf e}_2)\sin\Omega,
\end{equation*}
we can write
\begin{equation}
  \frac{\partial{L}}{\partial{{\bf r}}}=-\frac{\partial{\Omega}}{\partial{{\bf r}}}\cos I+
  \frac{1}{r}{{\bf e}}_{\nu}^T,\qquad
  \frac{\partial{L}}{\partial{\dot{{\bf r}}}}=-\frac{\partial{\Omega}}{\partial{\dot{{\bf r}}}}\cos I,
  \label{eq:dpsi_drv}
\end{equation}
where we have used
\begin{equation*}
  \frac{\partial{{{\bf e}}_r}}{\partial{{{\bf r}}}}=(I_d-{\bf e}_r{{\bf e}}_r^T),
\end{equation*}
and $I_d$ is the $3\times 3$ identity matrix.

The Euler parameters $\iota_5$, $\iota_6$, $\iota_7$, $\iota_8$ at the
initial time $t_*$ are written in terms of $L$, $\Omega$, $I$ by means
of Eqs.~(\ref{eq:q1234}), in which we set $\Psi=L$. Then, these
expressions are differentiated with respect to ${\bf r}$, $\dot{{\bf
    r}}$, and taking into
account~(\ref{eq:dOM_drv}),~(\ref{eq:dI_drv}),
and~(\ref{eq:dpsi_drv}), we obtain formulae~(\ref{eq:dq58_dr})
and~(\ref{eq:dq58_dv}).



\section{Partial derivatives of $\iota_1',\ldots,\iota_8'$ with respect to intermediate elements}
\label{sec:dfdi}
Let us recall that ${\bm\iota}=(\iota_1,\ldots,\iota_8)^T$. We define
\begin{multline*}
  {\cal K}_n=(rF_r-2\Up)\frac{\partial(ru_n)}{\partial{\bm\iota}}+
  ru_nF_r\frac{\partial r}{\partial{\bm\iota}}-\frac{1}{2}(\sigma P_r+hP_{\nu})
  \frac{\partial b_n}{\partial{\bm\iota}}\\
  +\frac{b_n}{4}\frac{\partial\iota'_3}{\partial{\bm\iota}}
  +ru_n\Bigl(r\frac{\partial F_r}{\partial{\bm\iota}}-
  2\frac{\partial\Up}{\partial{\bm\iota}}\Bigr),\quad n=1,2,4,5,
\end{multline*}
where
\begin{equation*}
  u_1=-U_1,\quad u_2=U_0,\quad u_4=U_2,\quad u_5=\frac{\iota_2U_1-\iota_1\iota_3U_2}{c\iota_1},
\end{equation*}
and
\begin{align*}
  b_1 & = -\iota_1\tilde{U}_2-\iota_2\tilde{U}_3-2\mu U_2^2,\\[2pt]
  b_2 & = \iota_1(2\chi+\tilde{U}_1)+\iota_2\tilde{U}_2+\mu(\tilde{U}_3-4U_3),\\[2pt]
  b_4 & = \iota_1(\tilde{U}_3-4U_3)+2\iota_2U_2^2+\mu(\tilde{U}_5-8U_5),\\[2pt]
  b_5 & = \frac{1}{2\iota_1}\Bigl[\frac{r}{c}(\iota_1U_1+\iota_2U_2)-cU_3\Bigl].
\end{align*}
The desired derivatives take the form
\begin{align}
  \frac{\partial\iota'_n}{\partial{\bm\iota}} & = {\cal K}_n,\quad n=1,2,4,\\[2pt]
  \frac{\partial\iota'_3}{\partial{\bm\iota}} & = -2\Bigl(P_r\frac{\partial\sigma}{\partial{\bm\iota}}
  +P_{\nu}\frac{\partial h}{\partial{\bm\iota}}+\sigma\frac{\partial P_r}{\partial{\bm\iota}}+
  h\frac{\partial P_{\nu}}{\partial{\bm\iota}}\Bigr),\\[2pt]
  2\frac{\partial\iota'_{k+4}}{\partial{\bm\iota}} & = \frac{\alpha_k}{r}\Bigl(\frac{\partial h}
  {\partial{\bm\iota}}-\frac{\partial c}{\partial{\bm\iota}}+\frac{c-h}{r}\frac{\partial r}
  {\partial{\bm\iota}}+r{\cal K}_5\Bigr)+\frac{r}{h}\Bigl(2F_z\frac{\partial r}{\partial{\bm\iota}}
  -F_z\frac{r}{h}\frac{\partial h}{\partial{\bm\iota}}\nonumber\\
  &\quad\,+r\frac{\partial F_z}{\partial{\bm\iota}}\Bigr)(\beta_kc_{\nu}+\gamma_ks_{\nu})
  +\frac{r^2}{h}F_z\Bigl[(\gamma_kc_{\nu}-\beta_ks_{\nu})\frac{\partial\nu}{\partial{\bm\iota}}
  +\frac{\partial\beta_k}{\partial{\bm\iota}}c_{\nu}+\frac{\partial\gamma_k}{\partial{\bm\iota}}s_{\nu}\Bigr]\nonumber\\
  &\quad\,+N\frac{\partial\alpha_k}{\partial{\bm\iota}},\quad k=1,2,3,4,
\end{align}
where $c_{\nu}$, $s_{\nu}$ denote $\cos\nu$, $\sin\nu$, respectively,
and $\alpha_k$, $\beta_k$, $\gamma_k$ denote the $k$-th component of
the vectors $\alpha$, $\beta$, $\gamma$ defined below:
\begin{equation*}
  \alpha=(-\iota_8,\,\iota_7,\,-\iota_6,\,\iota_5),\qquad
  \beta=(-\iota_6,\,\iota_5,\,\iota_8,\,-\iota_7),\qquad
  \gamma =(-\iota_7,\,-\iota_8,\,\iota_5,\,\iota_6).
\end{equation*}
The partial derivatives of $r$, $\sigma$, $c$, $h$, $\nu$ are reported
in Section~{\ref{sec:drv_di}. Moreover, we need the following
relations:
\begin{align*}
  \frac{\partial b_1}{\partial{\bm\iota}} & = -\Bigl(\tilde{U}_2,\,\tilde{U}_3,\,\iota_1\tilde{U}_4
  +\frac{3}{2}\iota_2\tilde{U}_5+4\mu U_2U_4-\chi(\iota_1\tilde{U}_3+\iota_2\tilde{U}_4
  +2\mu U_2U_3),\,{\bf 0}_5\Bigr),\\[2pt]
  \frac{\partial b_2}{\partial{\bm\iota}} & = \Bigl(2\chi+\tilde{U}_1,\,\tilde{U}_2,\,\frac{1}{2}
  \iota_1\tilde{U}_3+\iota_2\tilde{U}_4+\frac{3}{2}\mu(\tilde{U}_5-4U_5)+\chi(b_1-2\mu U_4),\,{\bf 0}_5\Bigr),\\[2pt]
  \frac{\partial b_4}{\partial{\bm\iota}} & = \Bigl(\tilde{U}_3-4U_3,\,2U_2^2,\,\frac{3}{2}\iota_1
  (\tilde{U}_5-4U_5)+4\iota_2U_2U_4+\frac{5}{2}\mu(\tilde{U}_7-8U_7)\\
  &\quad\, -\chi[\iota_1(\tilde{U}_4-2U_4)+2\iota_2U_2U_3+\mu(\tilde{U}_6-4U_6)],\,{\bf 0}_5\Bigr),\\[2pt]
  2\iota_1c\frac{\partial b_5}{\partial{\bm\iota}} & = (\iota_1U_1+\iota_2U_2)\Bigl(\frac{\partial r}
  {\partial{\bm\iota}}-\frac{r}{c}\frac{\partial c}{\partial{\bm\iota}}\Bigr)-cU_3\frac{\partial c}
  {\partial{\bm\iota}}+\frac{1}{2}\Bigl(\frac{2}{\iota_1}(c^2U_3-r\iota_2U_2),\,2rU_2,\\
  &\quad\, r(\iota_1U_3+2\iota_2U_4)-3c^2U_5+\chi[c^2U_4-r(\iota_1U_2+\iota_2U_3)],\,{\bf 0}_5\Bigr),\\[2pt]
  \frac{\partial u_5}{\partial{\bm\iota}} & = \frac{1}{c\iota_1}\Bigl(-\frac{\iota_2}{\iota_1}U_1,
  \,U_1,\,\frac{1}{2}[\iota_2U_3-\chi(\iota_1U_1+\iota_2U_2)],\,{\bf 0}_5\Bigr)
  +u_5c\frac{\partial c}{\partial{\bm\iota}},
\end{align*}
where ${\bf 0}_5\in\mathbb{R}^5$ is a row vector of null
entries. Assuming that $\Up$ depends only on ${\bf r}$, $t$ (see the
remark in Section~\ref{sec:Prop_form}), we have
\begin{equation}
  \frac{\partial\Up}{\partial{\bm\iota}}=
  \frac{\partial\Up}{\partial{{\bf r}}}\frac{\partial{{\bf r}}}{\partial{\bm\iota}}+
  \frac{\partial\Up}{\partial t}\frac{\partial t}{\partial{\bm\iota}}.\label{eq:dU_di}
\end{equation}
Let us denote by ${\bf y}$ either ${\bf F}$ or
${\bf P}$, and with $y_{\ell}$ the component of
${\bf y}({\bf r},\dot{{\bf r}},t)$ along one of the
directions associated to ${\bf e}_r$, ${\bf e}_{\nu}$,
${\bf e}_z$. Then, we can write
\begin{equation}
  \frac{\partial y_{\ell}}{\partial{\bm\iota}}={\bf e}_{\ell}^T\frac{\partial{{\bf y}}}{\partial{\bm\iota}}
  +{{\bf y}}^T\frac{\partial{{\bf e}}_{\ell}}{\partial{\bm\iota}},\label{eq:der_yell}
\end{equation}
where
\begin{equation}
  \frac{\partial{{\bf y}}}{\partial{\bm\iota}}=
  \frac{\partial{{\bf y}}}{\partial{{\bf r}}}\frac{\partial{{\bf r}}}{\partial{\bm\iota}}+
  \frac{\partial{{\bf y}}}{\partial{\dot{{\bf r}}}}\frac{\partial{\dot{{\bf r}}}}
  {\partial{\bm\iota}}+\frac{\partial{{\bf y}}}{\partial t}\frac{\partial t}{\partial{\bm\iota}}.
  \label{eq:der_y}
\end{equation}
The matrices $\partial{{\bf r}}/\partial{\bm\iota}$,
$\partial{\dot{{\bf r}}}/{\partial{\bm\iota}}$ are provided in
Section~\ref{sec:drv_di}, together with
$\partial{{\bf e}}_r/\partial{\bm\iota}$,
$\partial{{\bf e}}_{\nu}/\partial{\bm\iota}$, while
$\partial{{\bf e}}_z/\partial{\bm\iota}$ can be easily obtained
from the expression
\begin{equation}
  {\bf e}_z=(2\iota_6\iota_8+2\iota_5\iota_7,\,2\iota_7\iota_8-2\iota_5\iota_6,
  \,\iota_5^2-\iota_6^2-\iota_7^2+\iota_8^2)^T.
\end{equation}
Also note that
\begin{equation}
  \frac{\partial{{\bf F}}}{\partial{{\bf r}}}=
  \frac{\partial{{\bf P}}}{\partial{{\bf r}}}-\frac{\partial(\nabla\Up)}{\partial{{\bf r}}},\qquad
  \frac{\partial{{\bf F}}}{\partial t}=
  \frac{\partial{{\bf P}}}{\partial t}-\frac{\partial(\nabla\Up)}{\partial t}.
\end{equation}
Finally, we have
\begin{equation}
  \frac{\partial t}{\partial{\bm\iota}}=\Bigl(U_1,\,U_2,\,\frac{1}{2}[\iota_1U_3+2\iota_2U_4
  +3\mu U_5-\chi(\iota_1U_2+\iota_2U_3+\mu U_4)],\,1,\,0,\,0,\,0,\,0\Bigr).
\end{equation}

\section{Identities for the universal functions}
\label{sec:UnivForm}
We collect the identities for the universal functions introduced in
Eq.~(\ref{eq:Un}) that we used to derive some equations of this paper
\citep[see][sects.~4.5, 4.6]{Battin_1999}. For simplicity, we omit the
argument $\alpha$ in the universal functions. These formulae are:
\begin{gather*}
  U_n(\chi)+\alpha U_{n+2}(\chi)=\frac{\chi^n}{n!},\,\,\, n\in\mathbb{N},\\[2pt]  
  U_0(\chi)^2+\alpha U_1(\chi)^2=1,\\[2pt]
  U_1(\chi)^2-U_0(\chi)U_2(\chi)=U_2(\chi),\\[2pt]
  U_0(\chi)U_3(\chi)-U_1(\chi)U_2(\chi)=U_3(\chi)-\chi U_2(\chi),\\[2pt]
  U_1(\chi)U_3(\chi)-U_2(\chi)^2=2U_4(\chi)-\chi U_3(\chi),
\end{gather*}
the double argument identities:
\begin{align*}
  U_0(2\chi) & = U_0(\chi)^2-\alpha U_1(\chi)^2,\\[2pt]
  U_1(2\chi) & = 2U_0(\chi)U_1(\chi),\\[2pt]
  U_2(2\chi) & = 2U_1(\chi)^2,\\[2pt]
  U_3(2\chi) & = 2U_3(\chi)+2U_1(\chi)U_2(\chi),\\[2pt]
  U_5(2\chi) & = 2U_1(\chi)U_4(\chi)+\chi^2 U_3(\chi)+2U_5(\chi),
\end{align*}
and the differential relations:
\begin{gather*}
  \frac{\partial U_0}{\partial\chi}=-\alpha U_1,\qquad \frac{\partial U_m}{\partial\chi}=U_{m-1},
  \quad m\in\mathbb{N^+},\\[2pt]
  \frac{\partial U_n}{\partial\alpha}=\frac{1}{2}(nU_{n+2}-\chi U_{n+1}),\quad n\in\mathbb{N}.
\end{gather*}

\bibliographystyle{spbasic} 
\bibliography{mybib}

\end{document}